\documentclass[twocolumn]{aastex631}

\received{August 4, 2020}
\revised{June 14, 2021}
\accepted{June 15, 2021}
\submitjournal{Astrophysical Journal}

\shorttitle{Occurrence-weighted Median Planets Discovered by Transit Surveys}
\shortauthors{Schlaufman \& Halpern}

\begin{document}

\title{The Occurrence-weighted Median Planets Discovered by Transit
Surveys Orbiting Solar-type Stars and Their Implications for Planet
Formation and Evolution}

\correspondingauthor{Kevin C. Schlaufman}
\email{kschlaufman@jhu.edu}

\author[0000-0001-5761-6779]{Kevin C. Schlaufman}
\affiliation{Department of Physics and Astronomy \\
Johns Hopkins University \\
3400 N Charles St \\
Baltimore, MD 21218, USA}

\author{Noah D. Halpern}
\affiliation{Department of Physics and Astronomy \\
Johns Hopkins University \\
3400 N Charles St \\
Baltimore, MD 21218, USA}

\begin{abstract}

\noindent
Since planet occurrence and primordial atmospheric retention probability
increase with period, the occurrence-weighted median planets discovered
by transit surveys may bear little resemblance to the low-occurrence,
short-period planets sculpted by atmospheric escape ordinarily used
to calibrate mass--radius relations and planet formation models.
An occurrence-weighted mass--radius relation for the low-mass planets
discovered so far by transit surveys orbiting solar-type stars requires
both occurrence-weighted median Earth-mass and Neptune-mass planets
to have a few percent of their masses in hydrogen/helium (H/He)
atmospheres.  Unlike the Earth that finished forming long after the
protosolar nebula was dissipated, these occurrence-weighted median
Earth-mass planets must have formed early in their systems' histories.
The existence of significant H/He atmospheres around Earth-mass
planets confirms an important prediction of the core-accretion model
of planet formation.  It also implies core masses $M_{\text{c}}$ in
the range $2~M_{\oplus}\lesssim M_{\text{c}}\lesssim 8~M_{\oplus}$
that can retain their primordial atmospheres.  If atmospheric escape
is driven by photoevaporation due to extreme ultraviolet (EUV) flux,
then our observation requires a reduction in the fraction of incident
EUV flux converted into work usually assumed in photoevaporation models.
If atmospheric escape is core driven, then the occurrence-weighted median
Earth-mass planets must have large Bond albedos.  In contrast to Uranus
and Neptune that have at least 10\% of their masses in H/He atmospheres,
these occurrence-weighted median Neptune-mass planets are H/He poor.
The implication is that they experienced collisions or formed in much
shorter-lived and/or hotter parts of their parent protoplanetary disks
than Uranus and Neptune's formation location in the protosolar nebula.

\end{abstract}

\keywords{Exoplanet atmospheres(487) --- Exoplanet evolution(491) ---
Exoplanet formation(492) --- Exoplanets(498) --- Mini Neptunes(1063) ---
Super Earths(1655)}

\section{Introduction}

The observable properties of the host stars of transiting planets can
be used to infer both the masses and radii of their transiting planets.
The radius of a planet with a given mass can then be used as a diagnostic
of its composition and consequently its formation and evolution.
Studies of planet occurrence around both solar-type and late-type dwarfs
in the Kepler field as a function of orbital period $P$ and planet radius
$R_{\text{p}}$ have revealed that long-period, small-radius planets
are much more common outcomes of planet formation than short-period,
large-radius planets \citep[e.g.,][]{fre13,dre13,dre15,hsu19,hsu20}.
Though these long-period, small-radius planets are more representative
of the Galaxy's planet population, the number of planets discovered in
a photon-noise limited transit survey $N_{\text{p}}$ is proportional to
semimajor axis $a$ and $R_{\text{p}}$ as $a^{-5/2}\,R_{\text{p}}^{6}$
\citep{pep03}.  In addition to being relatively uncommon among transit
discoveries, at constant mass it is more difficult to infer masses
using the Doppler technique or transit-timing variations (TTVs) for
these high-occurrence, long-period, small-radius planets than for
low-occurrence, short-period, large-radius planets.  The amplitude of
the Doppler reflex velocity $K$ for a star hosting a planet with mass
$M_{\text{p}}$ is proportional to $M_{\text{p}}\,P^{-1/3}$, dropping
below $K = 3$ m s$^{-1}$ for a planet with $M_{\text{p}} = 10~M_{\oplus}$
at $P = 10$ days.  The signal-to-noise ratio S/N in a TTV measurement is
proportional to $M_{\text{p}}\,R_{\text{p}}^{3/2}\,P^{5/6}$ in general
and $M_{\text{p}}\,R_{\text{p}}^{3/2}\,P^{1/3}$ in the duration-limited
case appropriate for Kepler, K2, and the Transiting Exoplanet Survey
Satellite (TESS) continuous-viewing zones \citep{ste16,mil17}.  The net
result of these observational selection effects is that high-occurrence,
long-period, small-radius planets are underrepresented in the existing
sample of small planets with inferred masses.

The underrepresentation of high-occurrence, long-period, small-radius
planets in the existing sample of mass inferences is an obstacle to
the use of inferred exoplanet masses and radii to calibrate models of
planet formation and evolution.  Mass and radius inferences are most
constraining for low-mass planets, as masses and radii do not uniquely
define the compositions of more massive planets \citep[e.g.,][]{ada08}.
While mass and radius measurements for short-period planets are
powerful probes of planet evolution, the susceptibility of primordial
hydrogen/helium (H/He) atmospheres to atmospheric escape make it
more difficult to discern the properties of low-mass, short-period
planets immediately after the end of the planet formation process
\citep[e.g.,][]{you11,lop12,lop13,lop14,owe13,owe16,owe17,wu13,gin16,gin18}.

Despite these difficulties, the relationship between planet
masses and radii has been quantified with increasing statistical
sophistication for planets orbiting both solar-type and late-type dwarfs
\citep[e.g.,][]{lis11b,wu13,wei14,had14,wol16,mil17,che17,nin18,kan19,ulm19,nei20,ote20,tes21}.
While these studies acknowledged the underrepresentation of
high-occurrence, long-period, small-radius planets in their analysis
samples, their goals did not require correcting for it.  Many of
these studies conflated planets orbiting solar-type and late-type
dwarfs \citep{wu13,wei14,had14,wol16,che17,nin18,ulm19,ote20,tes21}
even though it has become clear that the outcome of planet formation
depends on host star mass \citep[e.g.,][]{lau04,ida05}.  A few of these
studies chose to focus either on Doppler \citep{nei20} or TTV mass
inferences \citep{had14}.  Several limited the range of planetary radii
considered, thereby ignoring low-mass, large-radius planets like those in
the Kepler-51 system \citep{wei14,had14,wol16,nin18}.  Some studied only
planets with precise mass and radius inferences, biasing their results
against low-density planets for which precise mass measurements are more
challenging \citep{ote20}.

In this paper, we assemble a sample of masses and radii for transiting
planets orbiting single solar-type stars with Doppler-, TTV-, or
joint Doppler/TTV-inferred masses consistent with $M_{\text{p}}
\leq 20~M_{\oplus}$ without regard to planetary radii.  We leverage
the known occurrence of planets as a function of period and radius to
derive occurrence-weighted mass--radius, radius--mass, mass--density,
and radius--density relations.  While weighting the analysis sample
by occurrence does not completely eliminate the biases of the transit,
Doppler, or TTV techniques, it does mitigate those effects using currently
available data.   The resulting relations provide a better view of
the usual outcome of planet formation in the Galaxy than the existing
mass--radius relations in the literature and should provide more accurate
constraints for theoretical models of planet formation.  We describe the
construction of our sample of masses and radii in Section 2.  We detail
in Section 3 our analysis procedure and a theoretical model that allows
us to interpret our results in the planet formation context.  We review
our results and their implications for planet formation and evolution
in Section 4.  We conclude by summarizing our findings in Section 5.

\section{Data}

We queried the NASA Exoplanet Archive \citep{ake13} on 14 June 2021
for all transiting exoplanets with Doppler data or TTVs.  We first
removed all circumbinary planets.  We then queried SIMBAD to find
the Gaia Data Release 2 (DR2) source ID and SIMBAD default name of
every host star.  We next used those Gaia DR2 source IDs to query
the ESA Gaia Archive for each host star's photometric and astrometric
properties \citep{gai16,gai18,are18,eva18,ham18,lur18,rie18,mar19}.
We dereddened each star's Gaia $G_{\text{BP}}-G_{\text{RP}}$ color
using the three-dimensional reddening map from \citet{lal18} assuming
Gaia DR2 extinction coefficients from \citet{cas18} and prior-informed
stellar distances from \citet{bai18}.  We retained only planets orbiting
solar-type stars with $0.58 < \left(G_{\text{BP}}-G_{\text{RP}}\right)_{0}
< 1.41$, roughly corresponding to dwarf stars with effective temperature
$T_{\text{eff}}$ in the range 4500 K $\lesssim T_{\text{eff}} \lesssim$
6500 K or equivalently dwarf stars with spectral types between F5V and K5V
\citep[e.g.,][]{pec13}\footnote{\url{https://www.pas.rochester.edu/~emamajek/EEM_dwarf_UBVIJHK_colors_Teff.txt}}.
We used dereddened Gaia $\left(G_{\text{BP}}-G_{\text{RP}}\right)_{0}$
color to focus on solar-type stars instead of inferred stellar masses
$M_{\ast}$ or $T_{\text{eff}}$ because for every exoplanet host star
in our initial sample high-precision Gaia DR2 photometry is available
and $\left(G_{\text{BP}}-G_{\text{RP}}\right)_{0}$ can be homogeneously
inferred.

We analyze masses inferred using both the Doppler and TTV techniques,
as there is no reason to believe that the two methods produce
systematically different mass inferences \citep[e.g.,][]{mil17}.
To ensure the homogeneity of the mass inferences in our analysis sample
to the greatest extent possible, we prefer mass estimates based on Kepler
TTVs from the large, homogeneous Kepler TTV analysis of \citet{had17} when
they have not been superseded by subsequent analyses.  If \citet{had17}
found that it was not possible to robustly infer the mass of a planet
with a previously published TTV-based mass estimate, then we excluded
that planet from our analysis sample.  We also excluded planets with
published TTV-based mass estimates from \citet{xie14} and \citet{had14}
that were not reproduced in \citet{had17}, as those mass inferences were
based on analytic relations from \citet{lit12} and therefore were forced
to make assumptions about the unknown free eccentricity distributions
in their samples.  Since the TTV-based mass estimates of \citet{had17}
and \citet{saa17} were calculated without regard to planet radius, we
updated the radii and radius uncertainties for the planets analyzed in
those papers based on Kepler DR25 transit depths \citep{tho18} and host
star radii inferences from \citet{bre18}, \citet{joh17}, or \citet{ber20}
in that order.  Because our focus is on planets with $M_{\text{p}}
\lesssim 20~M_{\oplus}$, we excluded all planets with masses that are
not consistent with $M_{\text{p}} \leq 20~M_{\oplus}$ at the 3-$\sigma$
level or with only mass upper limits.  Our final analysis sample includes
201 planets orbiting single solar-type stars: 133 planets with Doppler
mass inferences, 58 planets with TTV mass inferences, and 10 planets
with masses from joint Doppler/TTV analyses.

\begin{figure*}
\plottwo{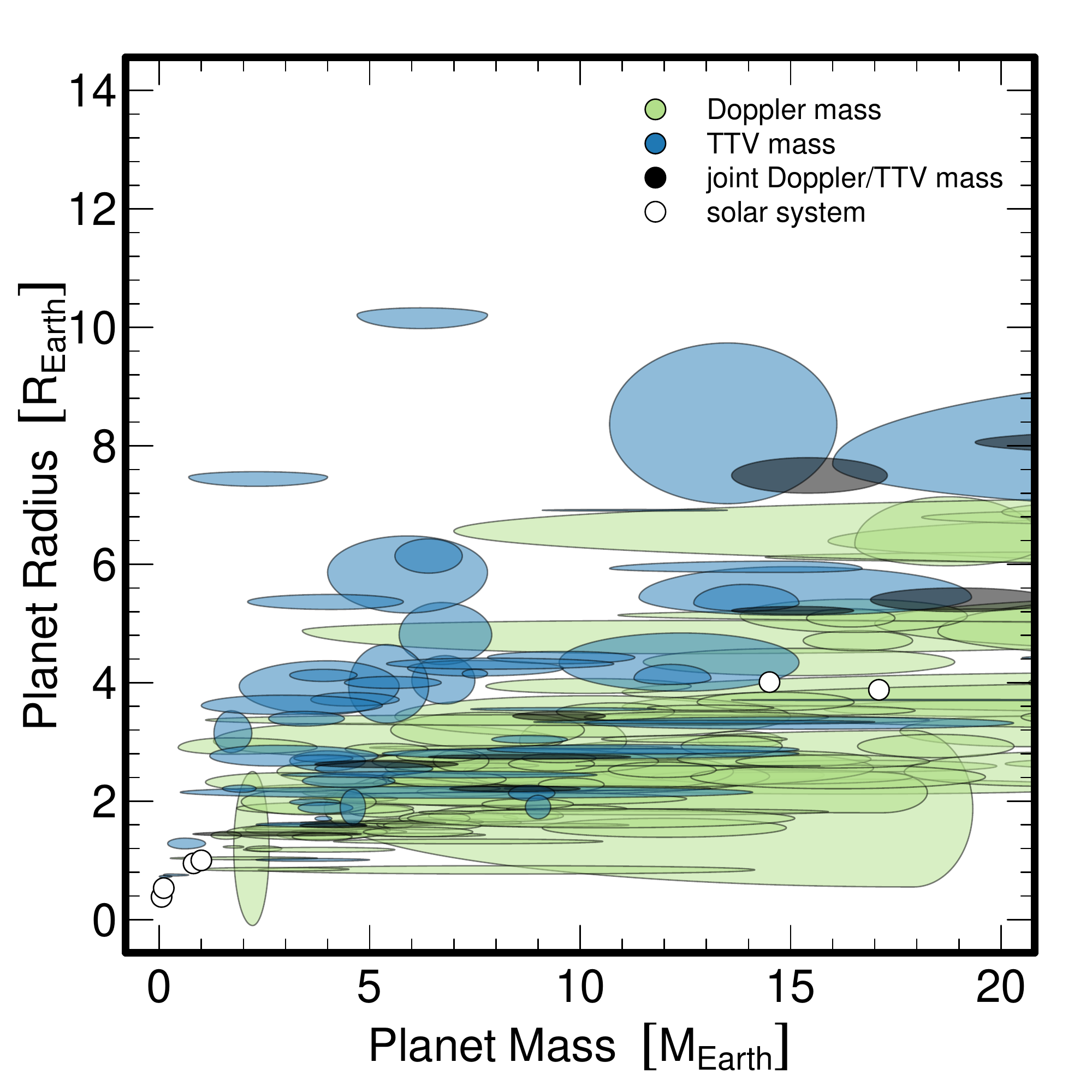}{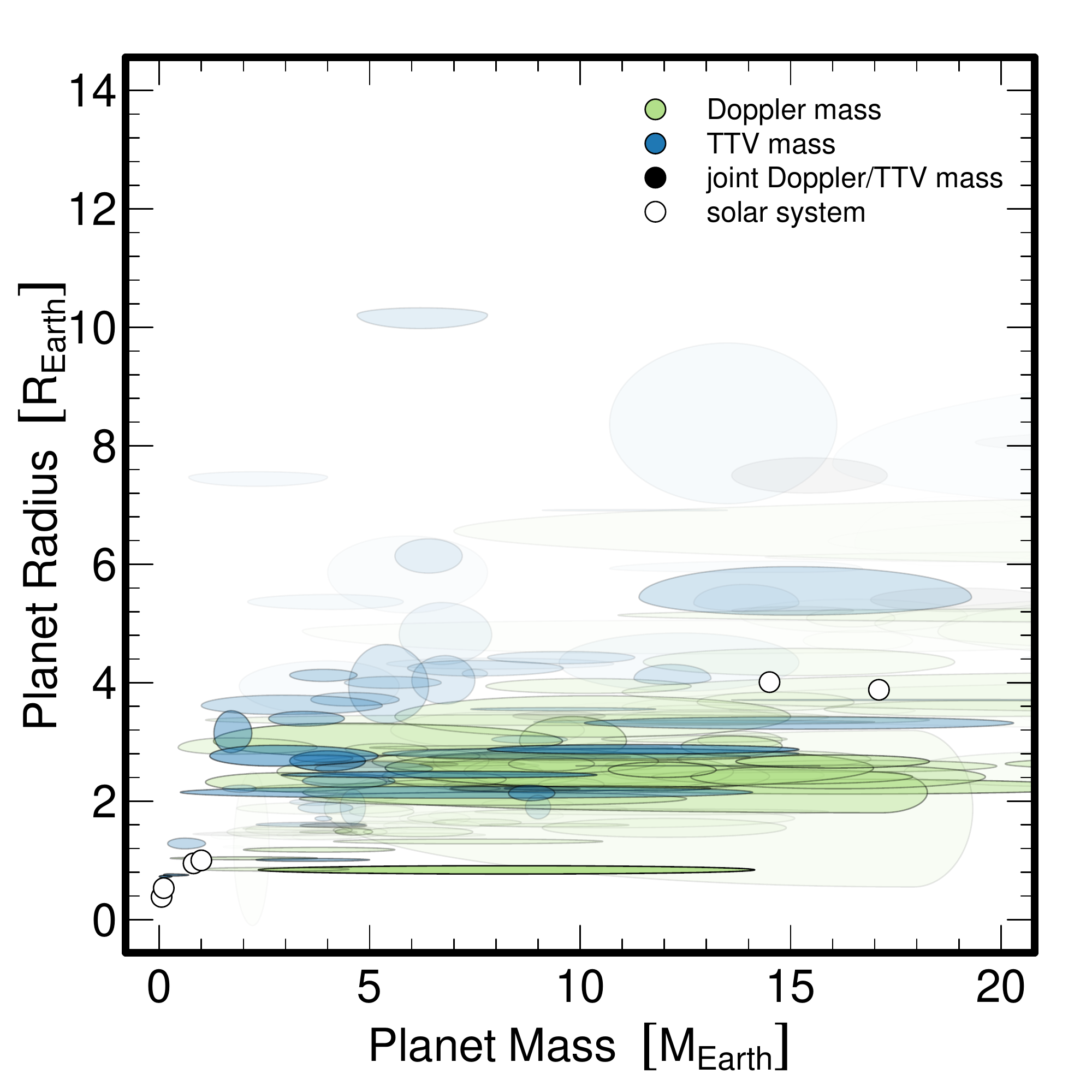}
\caption{Inferred masses and radii for planets with masses consistent
with $M_{\text{p}} \leq 20~M_{\oplus}$ at the 3-$\sigma$ level orbiting
solar-type stars listed in Table~\ref{tbl-1}.  Each ellipse is centered
on a planet's inferred mass and radius and represents the 1-$\sigma$
uncertainty in the mass and radius of the planet assuming no covariance
between mass and radius inferences.  We plot masses inferred by the
Doppler effect in blue, by transit-timing variations (TTVs) in green,
and by joint Doppler/TTV analyses in black.  We plot solar system
planets as open circles.  Left: opacity of each ellipse independent
of planet occurrence.  Right: opacity of each ellipse proportional
to the \citet{hsu19} occurrence of planets with the period and radius
of the planet represented by the ellipse.  The right panel provides a
less-biased view of the Galaxy's planet population than the heavily biased
view illustrated in the left panel.  Among the analysis sample of planets
orbiting solar-type stars with currently available mass inferences, the
most common planets have $2~R_{\oplus} \lesssim R_{\text{p}} \lesssim
3~R_{\oplus}$ over the mass range $1~M_{\oplus} \lesssim M_{\text{p}}
\lesssim 20~M_{\oplus}$.\label{fig01}}
\end{figure*}

We calculated the occurrence of each planet in our analysis sample by
linearly interpolating Table 2 of \citet{hsu19} that gives planet
occurrence as a function of period and radius when considering
both detection and vetting efficiency.  We plot our analysis sample
in Figure~\ref{fig01} and list each planet in the analysis sample
sorted in decreasing order of planet occurrence in Table~\ref{tbl-1}.
For multiple-planet systems, we order systems based on the maximum
occurrence among their individual planets.  Consequently, the planets
near the top of Table~\ref{tbl-1} are more frequent outcomes of
planet formation and evolution than the planets near the bottom.
We find that the occurrence-weighted mean period and insolation of
the planets in our analysis sample are $\overline{P} \approx 31$ days
and $\overline{F}_{\text{p}} \approx 27~F_{\oplus}$ respectively.  In
contrast, the mean period and insolation in our sample without occurrence
weighting are $\overline{P} \approx 21$ days and $\overline{F}_{\text{p}}
\approx 46~F_{\oplus}$ respectively.

\section{Analysis}

\begin{figure*}
\plottwo{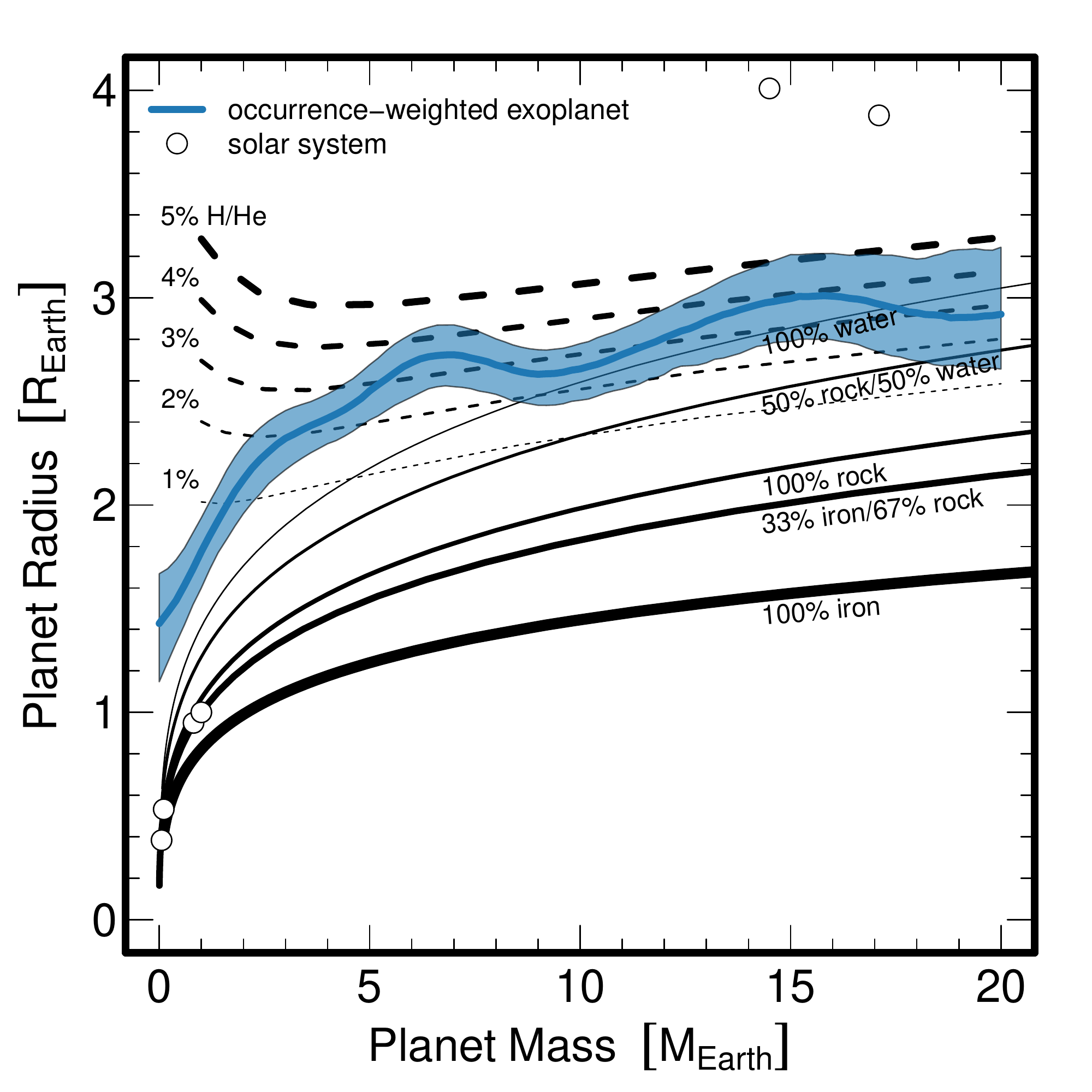}{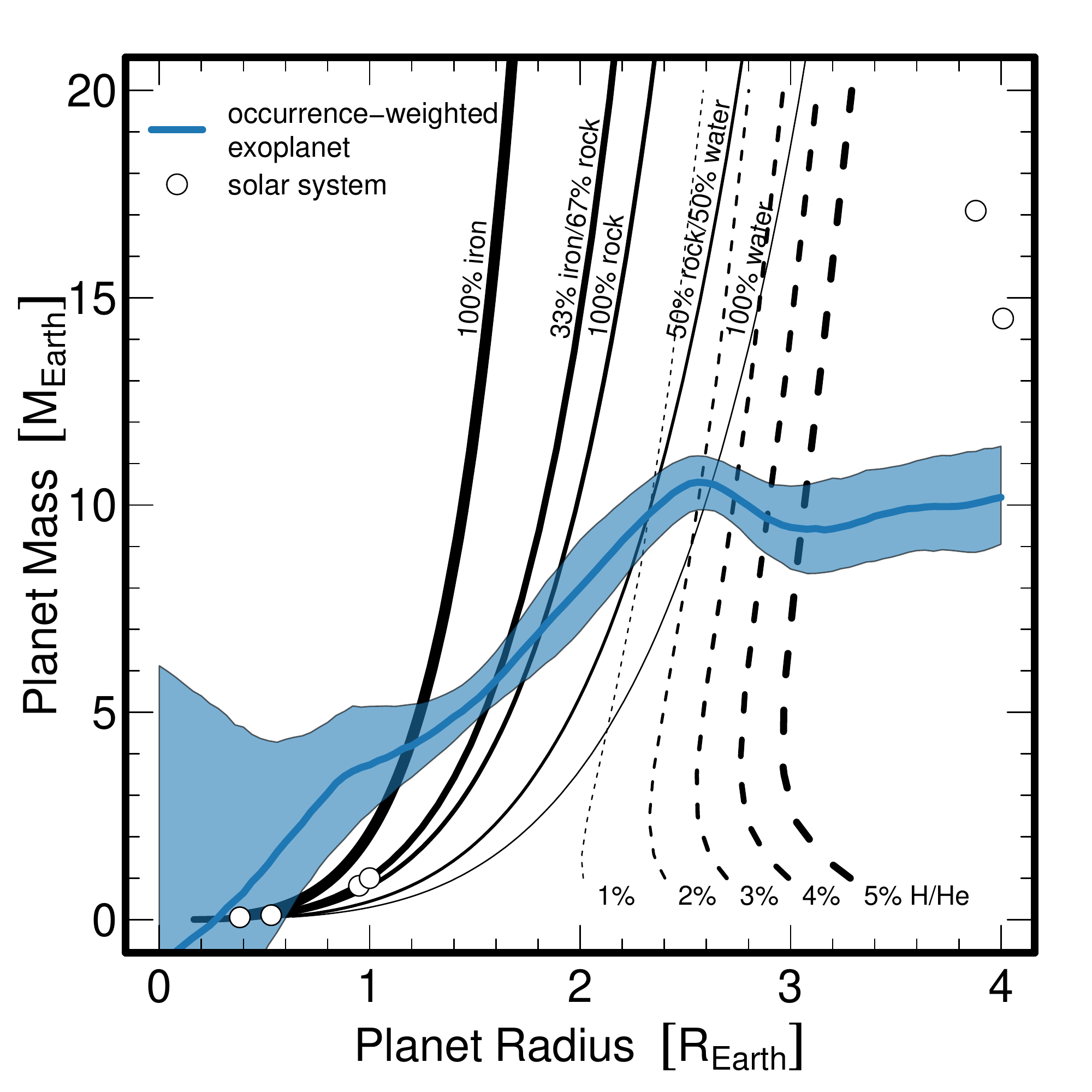}
\caption{Mass--radius and radius--mass relations resulting from
occurrence-weighted smoothing spline fits to individual iterations of
a Monte Carlo simulation sampling planet masses and radii from their
uncertainty distributions as described in Table~\ref{tbl-1}.  The solid
dark blue line corresponds to the median occurrence-weighted smoothing
spline radius or mass at given mass or radius.  The light blue region
spans the 16th to 84th percentiles of the occurrence-weighted smoothing
spline radius or mass at given mass or radius.  We plot solar system
planets as open circles.  We plot as solid black lines theoretical
models from \citet{zen19} for planets without H/He atmospheres and as
dashed black lines theoretical models from \citet{lop14} for planets
with significant H/He atmospheres.  For the \citet{lop14} models we
assume an age of 5 Gyr and insolation $F_{\text{p}} = 27~F_{\oplus}$
equal to the occurrence-weighted mean insolation of our sample.  Left:
occurrence-weighted exoplanet radius given mass.  Among the sample of
planets with currently available mass inferences, the occurrence-weighted
median planet at $M_{\text{p}} \approx 2.3~M_{\oplus}$ must have at least
1\% of its mass in a H/He atmosphere to explain its radius.  In contrast
to Uranus and Neptune that have at least 10\% of their masses in H/He
atmospheres, in the range $15~M_{\oplus} \lesssim M_{\text{p}} \lesssim
20~M_{\oplus}$ the occurrence-weighted median planet has at most 5\%
of its mass in a H/He atmosphere.  Right: occurrence-weighted exoplanet
mass given radius.  Among the sample of planets with currently available
mass inferences, the occurrence-weighted median planet with $R_{\text{p}}
\gtrsim 2.7~R_{\oplus}$ must have some H/He and the occurrence-weighted
median planet with $R_{\text{p}} \gtrsim 2.8~R_{\oplus}$ must have at
least 3\% of its mass in a H/He atmosphere.  The flat occurrence-weighted
radius--mass relation in the range $3~R_{\oplus} \lesssim R_{\text{p}}
\lesssim 4~R_{\oplus}$ shows that planets in that interval grow in
radius by accreting H/He and have a typical core mass $M_{\text{c}}
\approx 10~M_{\oplus}$.\label{fig02}}
\end{figure*}

\begin{figure*}
\plottwo{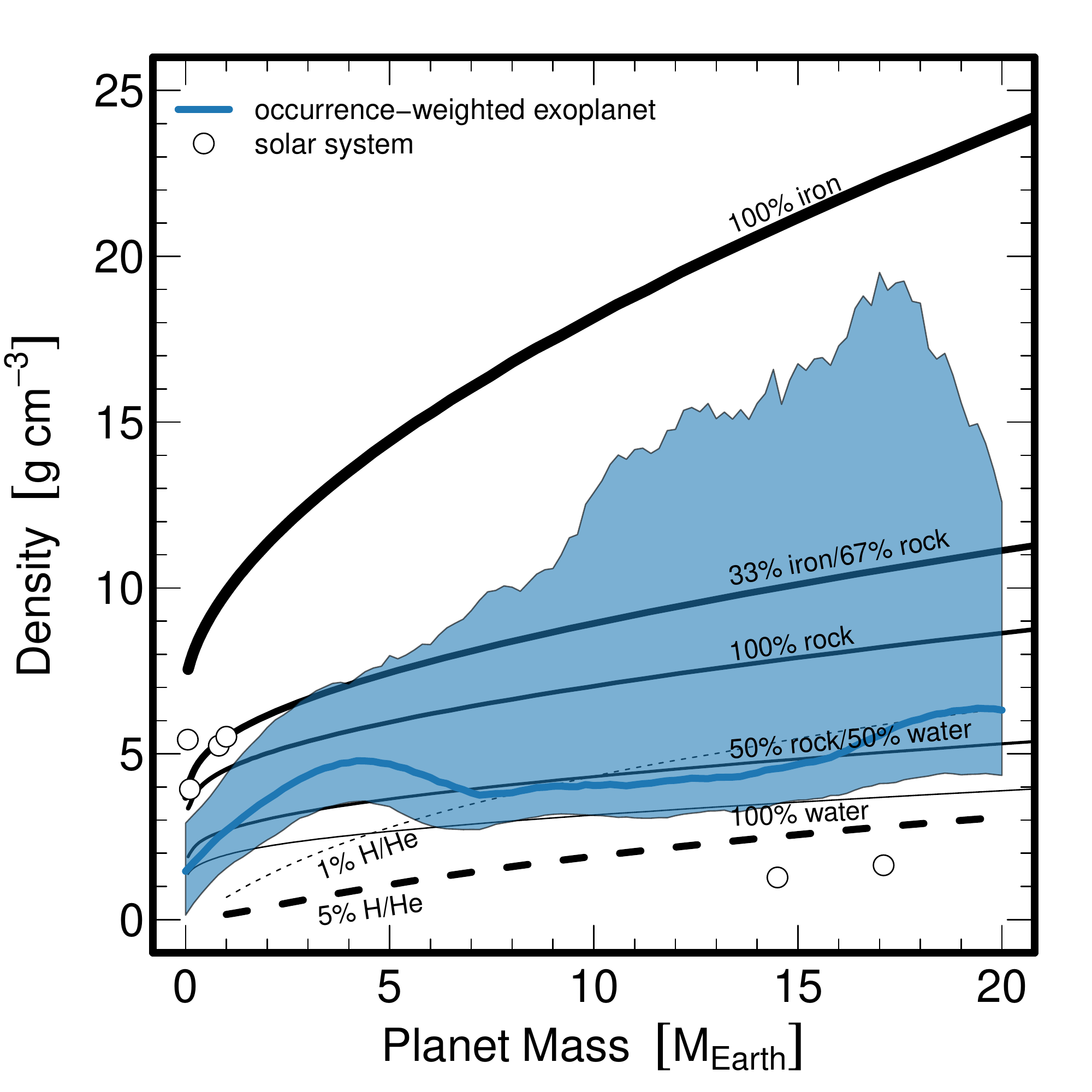}{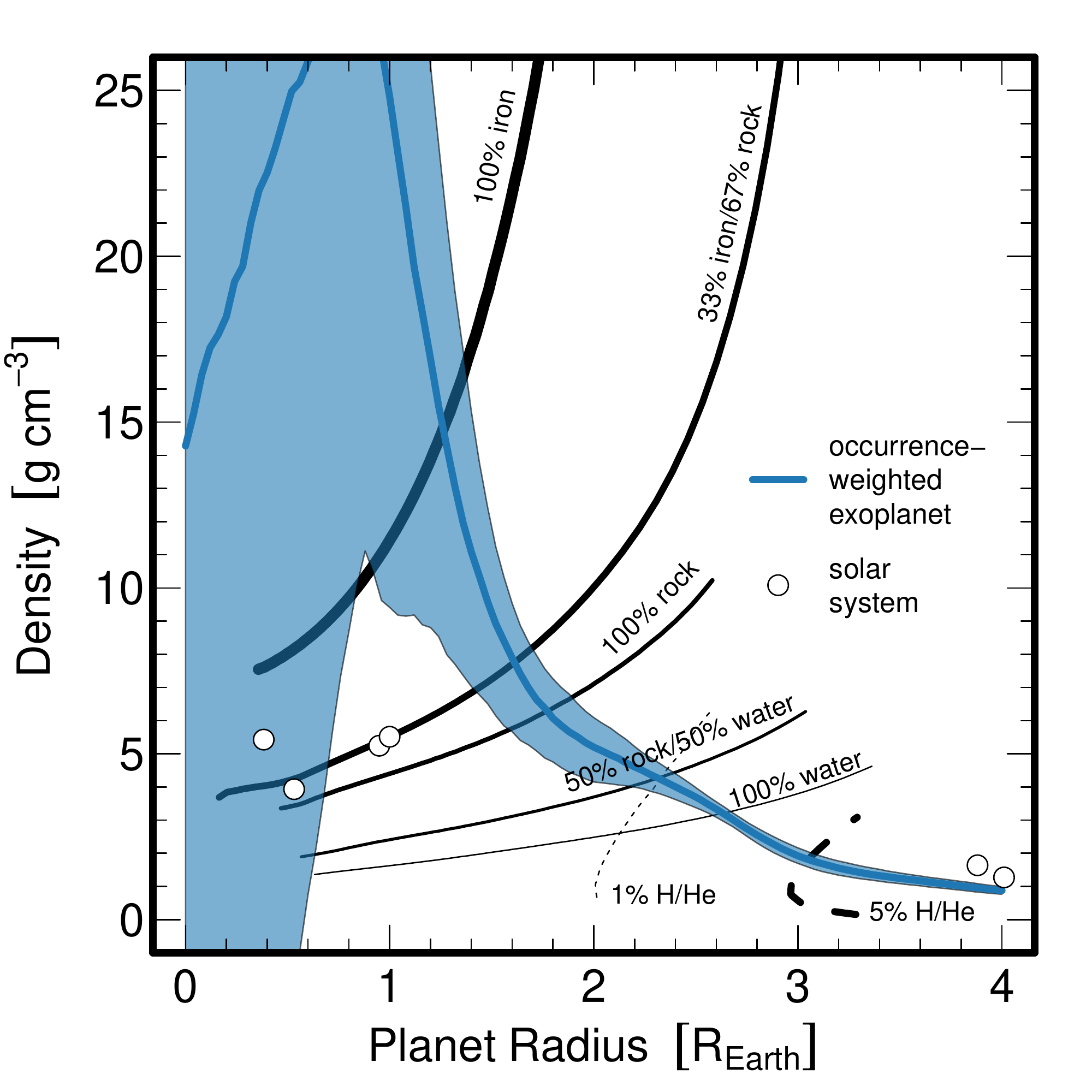}
\caption{Mass--density and radius--density relations resulting from
occurrence-weighted smoothing spline fits to individual iterations
of a Monte Carlo simulation sampling planet masses and radii from
their uncertainty distributions as described in Table~\ref{tbl-1}.
The solid dark blue line corresponds to the median occurrence-weighted
smoothing spline density at given mass or radius.  The light blue region
spans the 16th to 84th percentiles of the occurrence-weighted smoothing
spline density at given mass or radius.  We plot solar system planets
as open circles.  We plot theoretical models from \citet{zen19} for
planets without H/He atmospheres as black lines and theoretical models
from \citet{lop14} for planets with significant H/He atmospheres as
dashed lines.  For the \citet{lop14} models we assume an age of 5 Gyr and
insolation $F_{\text{p}} = 27~F_{\oplus}$ equal to the occurrence-weighted
mean insolation of our sample.  Left: occurrence-weighted exoplanet
density given mass.  Among the sample of planets with currently available
mass inferences, the occurrence-weighted median planet is less dense
than the terrestrial planets but denser than the ice giants in our solar
system.  Right: occurrence-weighted exoplanet density given radius.
Among the sample of planets with currently available mass inferences,
the occurrence-weighted median planet with $R_{\text{p}} \gtrsim
2.7~R_{\oplus}$ must have some H/He and the occurrence-weighted median
planet with $R_{\text{p}} \gtrsim 2.8~R_{\oplus}$ must have at least 3\%
of its mass in a H/He atmosphere.\label{fig03}}
\end{figure*}

We calculate occurrence-weighted median mass--radius, radius--mass,
mass--density, and radius--density relations and their uncertainties
using Monte Carlo simulations.  On each iteration, we sample masses
and radii for every planet in our analysis sample from the normal or
asymmetric normal distributions representing the published uncertainties
in masses and radii listed in Table~\ref{tbl-1}.  We assume that mass
and radius inferences are uncorrelated.  We next fit a cubic smoothing
spline using the \texttt{smooth.spline} function with smoothing parameter
$s = 0.9$ as implemented in \texttt{R} \citep{r20} while weighting each
planet's mass and radius point by the occurrence of that planet listed
in Table~\ref{tbl-1}.  We use the smoothing spline to predict radius as a
function of mass, mass as a function of radius, density as a function of
mass, and density as a function of radius at 101 points evenly separated
in the ranges $0~M_{\oplus} \leq M_{\text{p}} \leq 20~M_{\oplus}$ or
$0~R_{\oplus} \leq R_{\text{p}} \leq 4~R_{\oplus}$ and save the result.
We repeat this process 1,000 times to generate 1,000 occurrence-weighted
mass--radius, radius--mass, mass--density, and radius--density relations.
We illustrate the median occurrence-weighted mass--radius, radius--mass,
mass--density, and radius--density relations and their uncertainties in
Figures~\ref{fig02} and~\ref{fig03}.  We also include the full sample
of 1,000 occurrence-weighted mass--radius, radius--mass, mass--density,
and radius--density relations as machine-readable tables to enable
others to make use of our occurrence-weighted relations.  Characterizing
our occurrence-weighted sample with $\overline{F}_{\text{p}} \approx
27~F_{\oplus}$ as a single population with some dispersion as opposed
to multiple populations is justified in this case because the small
planet peak defining the so-called ``Fulton Gap'' disappears at planet
insolation $F_{\text{p}} \lesssim 30~F_{\oplus}$ \citep[e.g.,][]{ful17}.

We compare our occurrence-weighted relations to theoretical models
for both solid planets and rocky planets with significant H/He
atmospheres.  For solid planets, we use models from \citet{zen19}.
For rocky planets with significant H/He atmospheres, we interpolate
the grid of models presented in \citet{lop14} to calculate radii as
a function of mass and H/He atmosphere mass fraction for a 5 Gyr old
planet population experiencing the occurrence-weighted mean insolation
of our sample $\overline{F}_{\text{p}} = 27~F_{\oplus}$.  We find that
our occurrence-weighted mass--radius relation demands that planets
with $M_{\text{p}} \approx 2.3~M_{\oplus}$ must have at least 1\% of
their masses in H/He atmospheres.  Our occurrence-weighted mass--radius
relation indicates that in the range $0~M_{\oplus} \leq M_{\text{p}} \leq
7~M_{\oplus}$, the fractional mass in a planet's H/He atmosphere increases
with mass from 1\% at $M_{\text{p}} \approx 2~M_{\oplus}$ to more than
3\% at $M_{\text{p}} \approx 7~M_{\oplus}$.  Our occurrence-weighted
mass--radius relation provides no reason to believe that the fractional
mass in a planet's H/He atmosphere scales with mass in the range
$7~M_{\oplus} \lesssim M_{\text{p}} \lesssim 20~M_{\oplus}$.  Indeed, in
the mass range $9~M_{\oplus} \lesssim M_{\text{p}} \lesssim 20~M_{\oplus}$
the occurrence-weighted mass--radius relation is consistent with both
rocky planets with H/He atmospheres and planets made entirely of water
without H/He atmospheres.  In the mass range corresponding to the
ice giants Uranus and Neptune in our own solar system ($15~M_{\oplus}
\lesssim M_{\text{p}} \lesssim 20~M_{\oplus}$), the occurrence-weighted
median planet has at most 5\% of its mass in a H/He atmosphere.  This is
in contrast to Uranus and Neptune that have at least 10\% of their masses
in H/He atmospheres \citep[e.g.,][]{pod19,hel20}.

Our occurrence-weighted radius--mass relation requires that planets
with $R_{\text{p}} \gtrsim 2.7~R_{\oplus}$ must have some H/He and
planets with $R_{\text{p}} \gtrsim 2.8~R_{\oplus}$ must have at least
3\% of their masses in H/He atmospheres.  Below these critical radii,
the occurrence-weighted radius--mass relation cannot exclude planets made
entirely of water without H/He atmospheres.  The flat occurrence-weighted
radius--mass relation in the range $3~R_{\oplus} \lesssim R_{\text{p}}
\lesssim 4~R_{\oplus}$ implies that planets in that interval grow in
radius by accreting H/He without adding much mass.  This growth in
radius at constant mass also suggests a typical core mass $M_{\text{c}}
\approx 10~M_{\oplus}$.

Our occurrence-weighted mass--density and radius--density relations
are much less precise because density estimates are affected by both
mass and radius uncertainties.  Nevertheless, the occurrence-weighted
median $M_{\text{p}} \approx 1~M_{\oplus}$ planet is less dense than
the terrestrial planets in our solar system and the occurrence-weighted
median planet in the range $15~M_{\oplus} \lesssim M_{\text{p}} \lesssim
20~M_{\oplus}$ is more dense than the ice giants Uranus and Neptune.
The occurrence-weighted radius--density relation confirms that planets
with $R_{\text{p}} \gtrsim 2.7~R_{\oplus}$ must have some H/He and
planets with $R_{\text{p}} \gtrsim 2.8~R_{\oplus}$ must have at least 3\%
of their masses in H/He atmospheres.

\begin{figure*}
\plotone{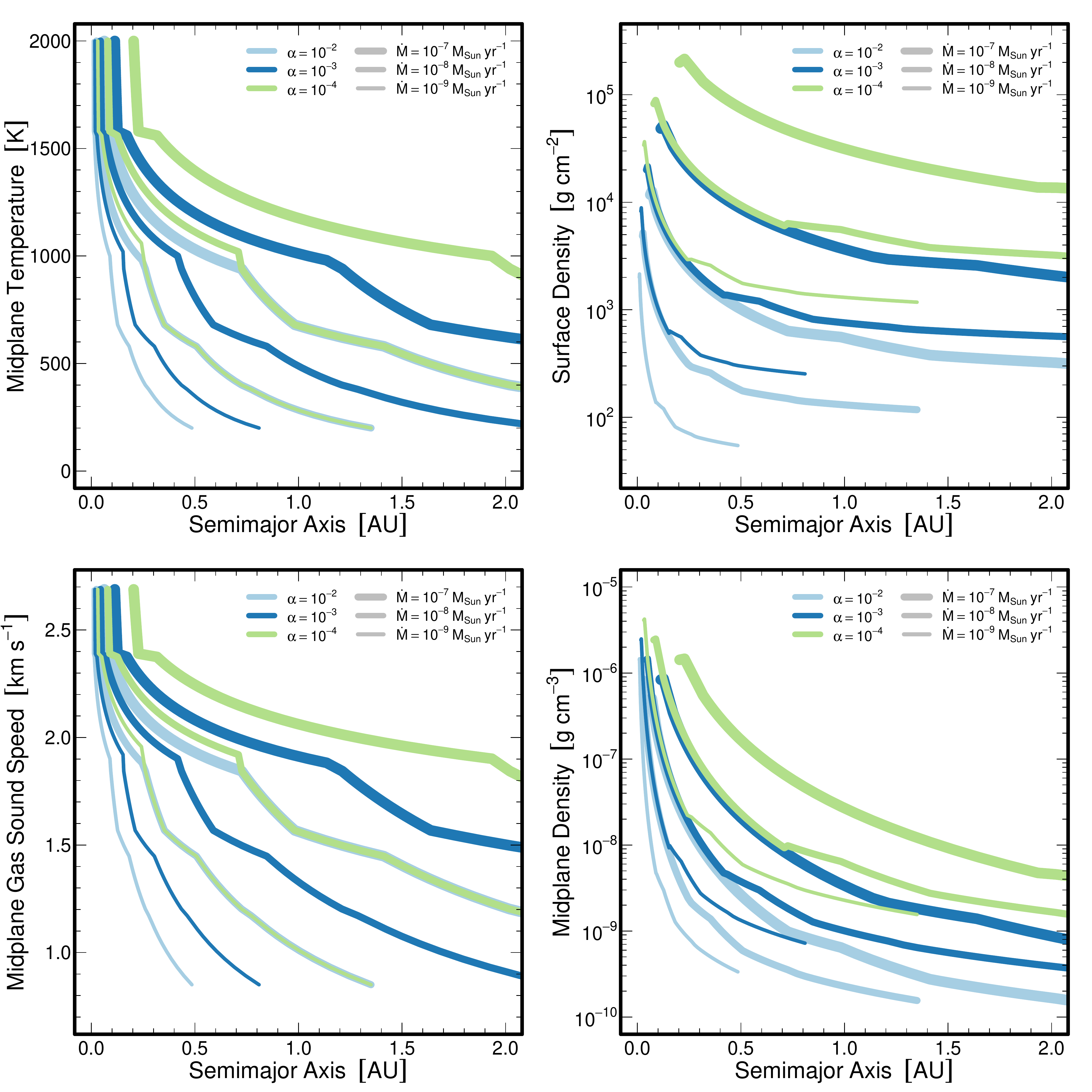}
\caption{Protoplanetary disk midplane temperature $T_{\text{disk}}$,
surface density $\Sigma$, midplane gas sound speed $c_{\text{s}}$,
and midplane density $\rho_{0}$ predicted by the analytic model of
\citet{bel97} as a function of Shakura-Sunyaev $\alpha$ parameter and
mass accretion rate.\label{fig04}}
\end{figure*}

\citet{arm10} derived the minimum planet mass necessary for a core to
maintain an isothermal H/He envelope with a small but non-negligible
mass fraction $\epsilon$
\begin{eqnarray}
M_{\text{p}} \gtrsim \left(\frac{3}{4\pi\rho_{\text{m}}}\right)^{1/2} \left(\frac{c_{\text{s}}^2}{G}\right)^{3/2} \left[\ln{\left(\frac{\epsilon\rho_{\text{m}}}{\rho_{\text{0}}}\right)}\right]^{3/2},\label{eq01}
\end{eqnarray}
where $\rho_{\text{m}}$ is the density of the core, $c_{\text{s}}$ and
$\rho_{\text{0}}$ are the gas sound speed and density in the midplane
of the core's parent protoplanetary disk, and $G$ is the gravitational
constant.  We use the analytic protoplanetary disk models of \citet{bel97}
to evaluate if our observation that the occurrence-weighted median
planet with $M_{\text{p}} \approx 2.3~M_{\oplus}$ must have at least
1\% of its mass in a H/He envelope is consistent with the expected
outcome of the planet formation process for a planet with $M_{\text{p}}
= 2.3~M_{\oplus}$.  We assume a mean molecular weight $\mu = 2.3$
appropriate for molecular hydrogen and helium in the solar ratio.  We
assume a $M_{\ast} = 1~M_{\odot}$ host star and calculate disk structure
as a function of Shakura-Sunyaev $\alpha$ parameter and mass accretion
rate $\dot{M}$.  We plot these disk models in Figure~\ref{fig04}.

\begin{figure*}
\plotone{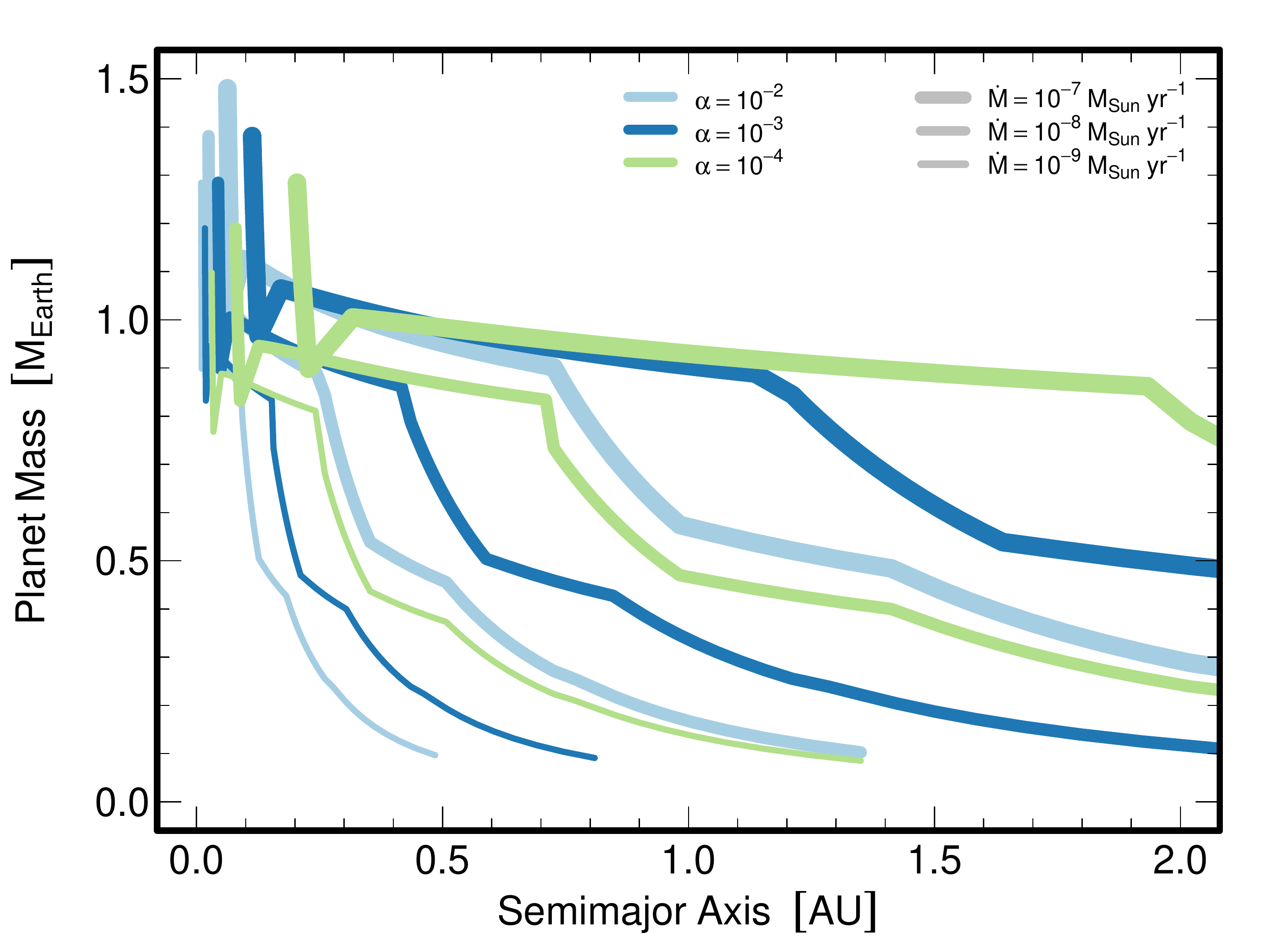}
\caption{Mass above which an embedded protoplanet can be expected
to retain an isothermal H/He envelope equal to 1\% of its total mass
accreted from its parent protoplanetary disk assuming the analytic disk
models of \citet{bel97} and an Earth-composition density.  The predictions
of this analytic disk model are in accord with our observation that the
occurrence-weighted median planet at $M_{\text{p}} \approx 2.3~M_{\oplus}$
has at least 1\% if its mass in a H/He atmosphere.\label{fig05}}
\end{figure*}

We use these disk models to estimate with Equation~(\ref{eq01}) the
mass necessary for a planet with density $\rho_{\text{m}} = 5.5$ g
cm$^{-3}$ consistent with the bulk density of the Earth to maintain an
isothermal H/He envelope with mass fraction $\epsilon = 0.01$.  We plot
the inferred minimum mass as a function of Shakura-Sunyaev $\alpha$
parameter and mass accretion rate in Figure~\ref{fig05}.  We find that
for Shakura-Sunyaev $\alpha$ parameter in the range $10^{-4} \leq \alpha
\leq 10^{-2}$ and mass accretion rate in the range $10^{-9}~M_{\odot}
\text{~yr}^{-1} \leq \dot{M} \leq 10^{-7}~M_{\odot} \text{~yr}^{-1}$,
planets with $M_{\text{p}} \gtrsim 1.5~M_{\oplus}$ should be able to
maintain isothermal H/He envelopes.  This theoretical expectation is fully
consistent with our observation that the occurrence-weighted median planet
with $M_{\text{p}} \approx 2.3~M_{\oplus}$ must have at least 1\% of its
mass in a H/He atmosphere to simultaneously explain its mass and radius.
While this is an idealized theoretical model, similar results are found
in state-of-the-art exoplanet population synthesis models with much more
detailed and comprehensive treatments of the physics of planet formation
\citep[e.g.,][]{ems20a,ems20b}.

\section{Discussion}

\begin{figure*}
\plotone{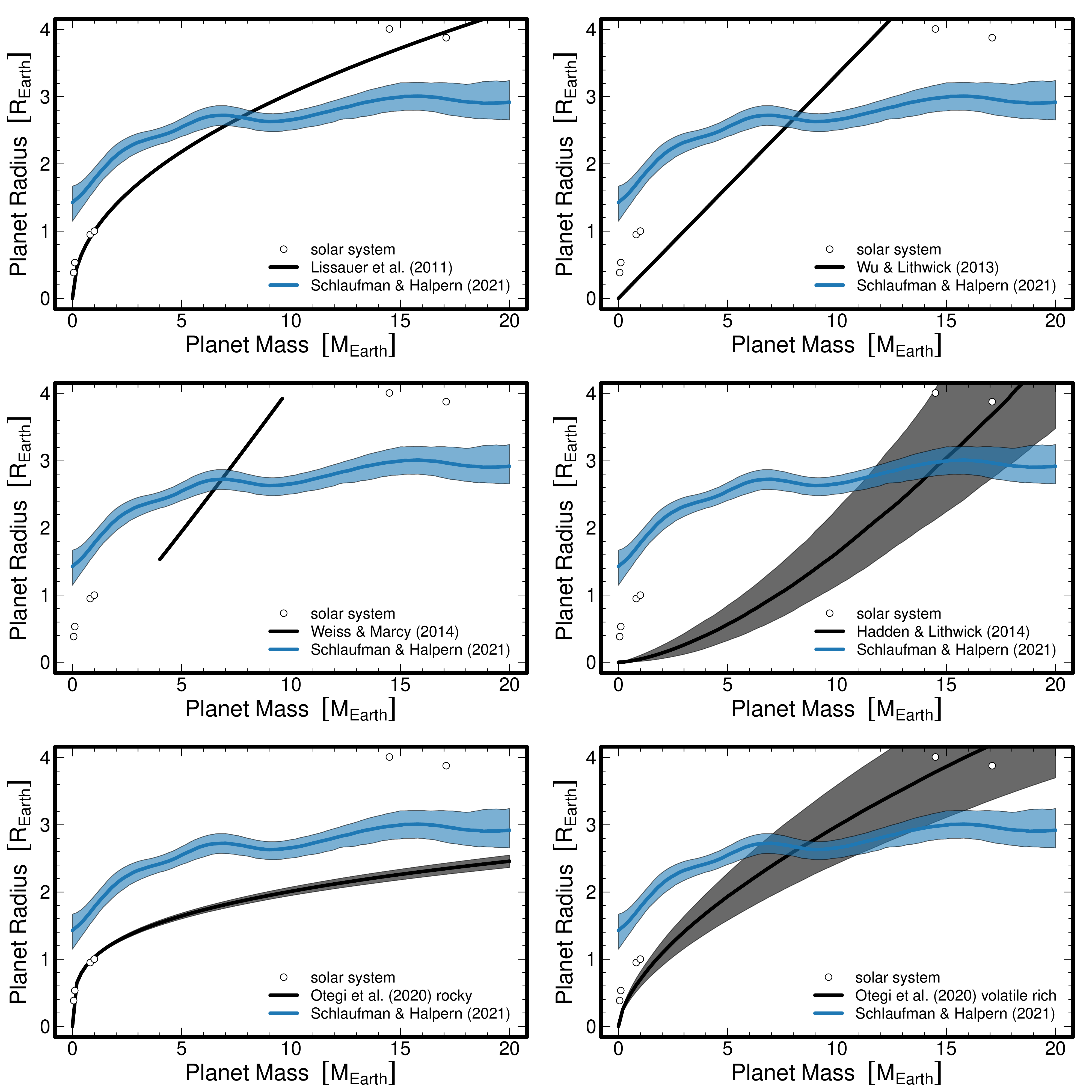}
\caption{Comparison of our occurrence-weighted mass--radius with
previously published relations without occurrence weighting.  We plot the
central value of each relation with a line and indicate its 1-$\sigma$
uncertainty if available with a shaded polygon.  Our occurrence-weighted
mass--radius relation indicates that planets with $M_{\text{p}} \approx
1~M_{\oplus}$ are larger and $M_{\text{p}} \approx 20~M_{\oplus}$ are
smaller than previously found.\label{fig06}}
\end{figure*}

\begin{figure*}
\plotone{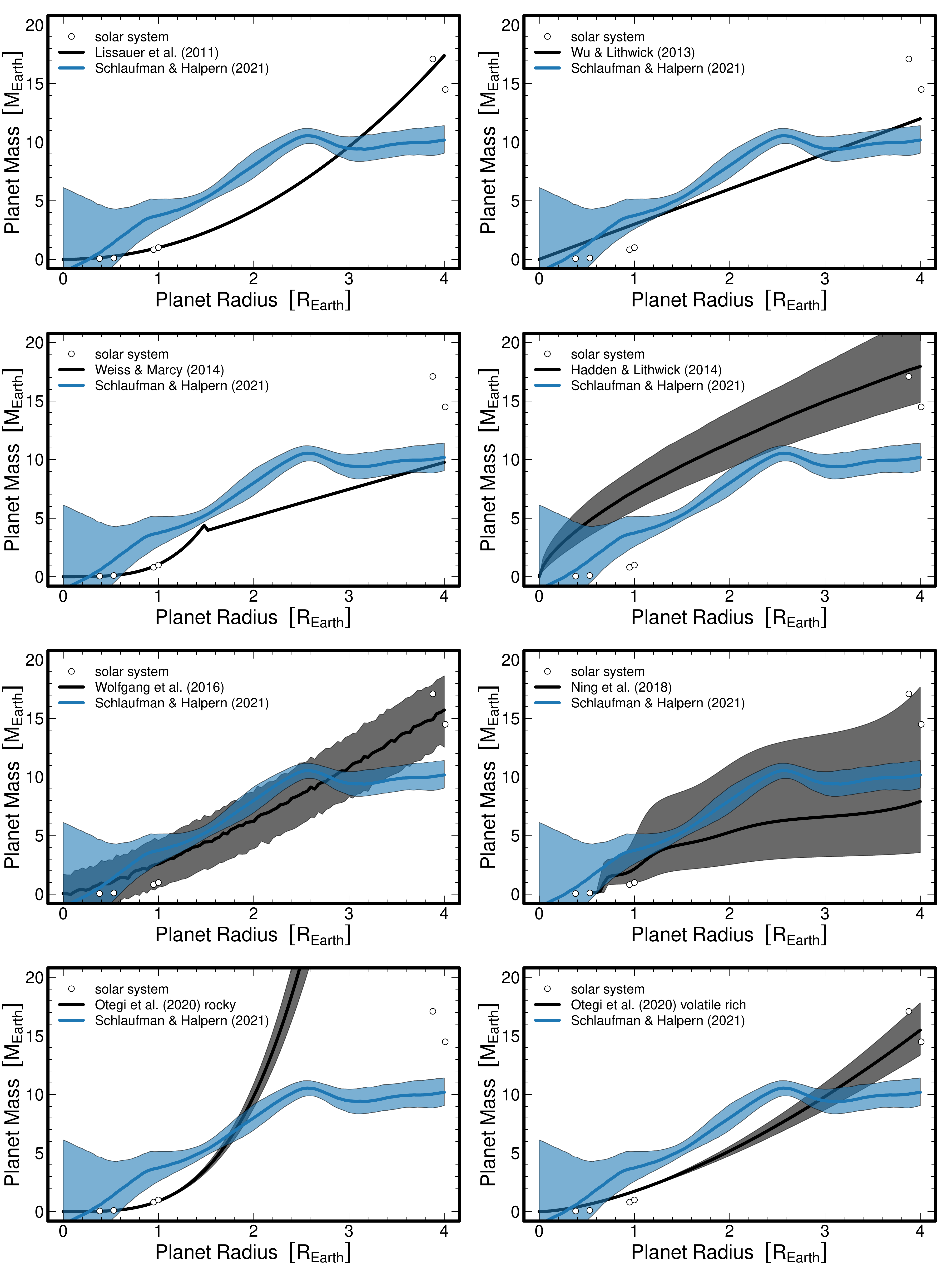}
\caption{Comparison of our occurrence-weighted mass--radius and
radius--mass relations with previously published relations without
occurrence weighting.  We plot the central value of each relation with a
line and indicate its 1-$\sigma$ uncertainty if available with a shaded
polygon.  Our occurrence-weighted radius--mass relation is consistent
with most relations in the literature.\label{fig07}}
\end{figure*}

We compare our occurrence-weighted mass--radius and radius--mass relations
to published mass--radius and radius--mass relations without occurrence
weighting in Figures \ref{fig06} and \ref{fig07}.  When compared to
the mass--radius relations in the literature, our occurrence-weighted
mass--radius relation suggests larger-radius $M_{\text{p}} \approx
1~M_{\oplus}$ planets and smaller-radius $M_{\text{p}} \approx
20~M_{\oplus}$ planets.  Our occurrence-weighted radius--mass relation
is consistent with most published radius--mass relations.

We find that the occurrence-weighted median planets discovered so far
by transit surveys have at least 1\% of their mass in H/He atmospheres
at $M_{\text{p}} \approx 2.3~M_{\oplus}$.  The fraction of mass in H/He
atmospheres increases from 1\% at $M_{\text{p}} \approx 2~M_{\oplus}$
to at least 3\% at $M_{\text{p}} \approx 7~M_{\oplus}$ then stays
consistent with 3\% up to $M_{\text{p}} \approx 20~M_{\oplus}$.
The occurrence-weighted median planet is also consistent with a 100\%
water composition in the range $9~M_{\oplus} \lesssim M_{\text{p}}
\lesssim 20~M_{\oplus}$.  The occurrence-weighted median planets
discovered so far by transit surveys are at constant mass less dense
than the terrestrial planets and more dense than the ice giants.

A planet with a significant H/He atmosphere must have formed before
the dissipation of its parent protoplanetary disk.  Observations of
the fraction of stars in young clusters with disks or accretion from
disks as a function of cluster age indicate that disk dissipation
occurs on a characteristic timescale $t_{\text{disk}} \approx 3$
Myr \citep[e.g.,][]{hai01,cie07,fed10}.  The short characteristic
disk dissipation time therefore requires the formation of the
occurrence-weighted median terrestrial-mass planets early in their host
systems' evolution.  This early formation is in contrast to the Earth,
which only reached $M_{\text{p}} \approx 0.6~M_{\oplus}$ in the first
4.5 Myr after the formation of calcium \& aluminum rich inclusions (CAIs)
at the dawn of the solar system and did not finish forming its core for
an additional 40 Myr \citep[e.g.,][]{kle09,lam21}.  It may be that the
solar system lacks short-period super-Earth or mini-Neptune planets
like those found by Kepler because of differences in the timescale
for planetary embryos to reach $M_{\text{p}} \approx 1~M_{\oplus}$.
In planetary systems with short-period super-Earth or mini-Neptune
planets, planetary embryos must have grown to $M_{\text{p}} \gtrsim
1~M_{\oplus}$ before the gas-rich phase of their parent protoplanetary
disks' evolution ended.  On the other hand, it appears that the planetary
embryos in the inner solar system that would grow into the terrestrial
planets did not reach $M_{\text{p}} \approx 1~M_{\oplus}$ during the
gas-rich phase of the protosolar nebula's evolution even though models
suggest that oligarchic growth at 1 AU should have been complete 1 Myr
after the formation of planetesimals \citep[e.g.,][]{kok00}.

The fact that the occurrence-weighted median planet has no more than
a few percent of its mass in a H/He envelope over the mass range
$2~M_{\oplus} \lesssim M_{\text{p}} \lesssim 8~M_{\oplus}$ requires a
core mass distribution extending from $M_{\text{c}} = 2~M_{\oplus}$
to at least $M_{\text{c}} = 8~M_{\oplus}$.  Parameterized models of
photoevaporative mass loss reach the same conclusion in a very different
way \citep[e.g.,][]{rog21}, supporting the robustness of this inference.
Similar core mass distributions can be realized by planet formation in
gas-poor protoplanetary disks \citep[e.g.,][]{lee21}, though a mixture
or rocky and icy cores may be necessary simultaneously explain both low-
and high-mass extremes of the distribution \citep[e.g.,][]{ven20a,ven20b}.

The existence of a 1\% H/He atmosphere around a planet with $M_{\text{p}}
\approx 2~M_{\oplus}$ orbiting a solar-type star with $F_{\text{p}}
\approx 27~F_{\oplus}$ can be used to constrain models of atmospheric
escape.  \citet{lop13} simulated the photoevaporation of low mean
molecular weight atmospheres of low-mass planets by stellar extreme
ultraviolet (EUV) irradiation as a function core mass, atmosphere
mass, system age, and efficiency of mass loss.  According to those
simulations, our observation of a 1\% H/He atmospheres around a planet
with $M_{\text{p}} \approx 2~M_{\oplus}$ several Gyr old can only be
accommodated if the fraction of incident EUV flux converted into work
is relatively low $\epsilon_{\text{evap}} \sim 0.01$.  This is an order
of magnitude lower than the value $\epsilon_{\text{evap}} \sim 0.1$
inferred from detailed models of EUV-driven mass loss from hot Jupiters,
short-period Neptune-mass planets, and the Kepler-11 and Kepler-36 systems
\citep[e.g.,][]{mur09,owe12,lop12,lop13}.  While it has usually been
assumed that a solar-type star's X-ray and EUV luminosities decay at the
same rate, \citet{kin21} have recently suggested that EUV emission falls
off much more slowly than X-ray emission.  As a consequence, a planet
experiences most of its total total X-ray and EUV irradation between
100 Myr and 10 Gyr after its formation.  This extended irradiation may
be supported by the observational results of \citet{dav21} and could
complicate the interpretation of the existence of 1\% H/He atmosphere
around a planet with $M_{\text{p}} \approx 2~M_{\oplus}$ orbiting a
solar-type star in the context of the \citet{lop13} model.

The alternative core-driven mass loss model proposed by \citet{gin16}
can only accommodate the existence of 1\% H/He atmospheres around
planets with $M_{\text{p}} \approx 2~M_{\oplus}$ if those planets have
low equilibrium temperatures and therefore compact atmospheres.  At the
occurrence-weighted average semimajor axis $\overline{a} = 0.19$ AU of
our analysis sample, an equilibrium temperature that low can only be
achieved if the occurrence-weighted median terrestrial-mass planets have
Bond albedos $A_{B} \gtrsim 0.95$.  Regardless of the mass loss mechanism,
we predict that more planets with $M_{\text{p}} \sim 1~M_{\oplus}$ and
$R_{\text{p}} \approx 2~R_{\oplus}$ orbiting solar-type stars will be
discovered and characterized as mass and radius inferences become possible
for low-mass planets with $P \gtrsim 40$ days, $a \gtrsim 0.25$ AU,
and $F_{\text{p}} \lesssim 20~F_{\oplus}$ \citep[e.g.,][]{ale20,nei20}.

The occurrence-weighted median planets in the mass range $15~M_{\oplus}
\lesssim M_{\text{p}} \lesssim 20~M_{\oplus}$ are massive enough
to avoid the loss of their atmospheres due to EUV or core-driven
mass loss, so the observation that they have no more than than 5\%
of their masses in H/He atmospheres constrains their formation.
Hot disks produce small H/He atmospheres at constant core mass
\citep{iko12}.  Short-lived disks and large opacities (possibly
maintained by atmospheric recycling with disk gas) may also keep
H/He mass fractions low \citep[e.g.,][]{lee15,gin16,ogi20,orm21}.
Exoplanet population synthesis calculations can produce Neptune-mass
planets with small H/He atmospheres if collisions with other planet
mass bodies occur before the parent protoplanetary disk is completely
dissipated \citep[e.g.,][]{ems20a,ems20b,sch21}.  These collisions could
be a ubiquitous feature of planet formation as systems of planets driven
into mean-motion resonances by disk migration may experience dynamical
instabilities and giant impacts as their parent protoplanetary disk
dissipates \citep[e.g.,][]{izi17,ogi20}.

The left-hand panel of Figure \ref{fig02} further shows that the
occurrence-weighted median planet in the mass range $10~M_{\oplus}
\lesssim M_{\text{p}} \lesssim 20~M_{\oplus}$ is core dominated, either
icy or rocky with a small H/He envelope.  The right-hand panel of Figure
\ref{fig02} shows that the occurrence-weighted median planet in the radius
range $3~R_{\oplus} \lesssim R_{\text{p}} \lesssim 4~R_{\oplus}$ grows
in radius but not mass by adding mass in a H/He atmosphere.  These two
observations suggest that the occurrence-weighted median planets with
$2~R_{\oplus} \lesssim R_{\text{p}} \lesssim 3~R_{\oplus}$ have a broad
range of core masses, while the occurrence-weighted median planets
with $R_{\text{p}} \gtrsim 3~R_{\oplus}$ have cores with $M_{\text{c}}
\approx 10~M_{\oplus}$.  This inference is in accord with the classical
threshold of $M_{\text{c}} \approx 10~M_{\oplus}$ for runaway envelope
accretion and giant planet formation \citep[e.g.,][]{pol96,hub05}.
We therefore argue that the occurrence-weighted median planets in the
mass range $15~M_{\oplus} \lesssim M_{\text{p}} \lesssim 20~M_{\oplus}$
formed (1) in hotter parts of their parent protoplanetary disks than
Uranus and Neptune, (2) shortly before the dissipation of their parent
protoplanetary disks, and/or (3) after experiencing collisions with other
planet-mass bodies as their parent protoplanetary disk was dissipated.
We favor the collision hypothesis for planets with $10~M_{\oplus}
\lesssim M_{\text{p}} \lesssim 20~M_{\oplus}$ and $2~R_{\oplus} \lesssim
R_{\text{p}} \lesssim 3~R_{\oplus}$.

While our occurrence-weighted mass--radius, radius--mass, mass--density,
and radius--density relations mitigate some of the observational
biases in previous studies, they do not completely eliminate them.
If at $P \lesssim 1$ yr Earth-composition planets with $M_{\text{p}}
\approx 2.3~M_{\oplus}$ and $R_{\text{p}} \approx 1.3~R_{\oplus}$
are a more common outcome of the planet formation process than H/He
atmosphere planets with $M_{\text{p}} \approx 2.3~M_{\oplus}$ and
$R_{\text{p}} \approx 2.2~R_{\oplus}$, then the former population would
be underrepresented even in data from the Kepler prime mission due to
the strong planet radius dependence of the number of detections in a
photon-noise limited transit survey.  Though the two populations would
have identical Doppler semiamplitudes, the population of $M_{\text{p}}
\approx 2.3~M_{\oplus}$ and $R_{\text{p}} \approx 1.3~R_{\oplus}$ planets
would often evade detection and therefore never be subject to follow-up
Doppler observations.  TTV-based mass inferences for the $M_{\text{p}}
\approx 2.3~M_{\oplus}$ and $R_{\text{p}} \approx 1.3~R_{\oplus}$
population would be more difficult as well.  The net result is that
there could be a population of Earth-composition, $M_{\text{p}} \approx
2.3~M_{\oplus}$, and $R_{\text{p}} \approx 1.3~R_{\oplus}$ planets that
finished forming after the end of the gas-rich phase of their parent
protoplanetary disks' evolution and is a more common outcome of planet
formation than the occurrence-weighted median planets described in
this paper.  Even if this so, then the conclusion that H/He atmosphere
planets with $M_{\text{p}} \approx 2.3~M_{\oplus}$ and $R_{\text{p}}
\approx 2.2~R_{\oplus}$ form before the end of the gas-rich phase of their
parent protoplanetary disks' evolution is unchanged.  The existence of a
largely unobserved population of Earth-composition, $M_{\text{p}} \approx
2.3~M_{\oplus}$, and $R_{\text{p}} \approx 1.3~R_{\oplus}$ planets at
$P \gtrsim \overline{P} = 31$ days does not affect our inferences about
photoevaporation, core-driven mass loss, or the structure \& formation
of the occurrence-weighted median Neptune-mass planets.

\section{Conclusion}

We derive occurrence-weighted mass--radius, radius--mass, mass--density,
and radius--density relations for the exoplanets discovered so far
by transit surveys.  These occurrence-weighted relations mitigate
some of the biases of the transit technique and the Doppler/TTV mass
inference methods.  They should provide a less-biased view of the typical
outcome of the planet formation process in the solar neighborhood than
mass--radius relations that have not accounted for planet occurrence.
We find that the occurrence-weighted median planet with $M_{\text{p}}
\approx 2.3~M_{\oplus}$ must have at least 1\% of its mass in a H/He
atmosphere.  The fraction of mass in the occurrence-weighted median
planet's H/He atmosphere increases from 1\% at $M_{\text{p}} \approx
2~M_{\oplus}$ to at least 3\% at $M_{\text{p}} \approx 7~M_{\oplus}$.
Our observation that the occurrence-weighted median planet has only a few
percent of its mass in a H/He envelope implies a core mass distribution
extending from $M_{\text{c}} = 2~M_{\oplus}$ to at least $M_{\text{c}}
= 8~M_{\oplus}$.  There is no discernible increase in H/He atmosphere
mass fraction in the range $7~M_{\oplus} \lesssim M_{\text{p}} \lesssim
20~M_{\oplus}$ and 100\% water compositions cannot be rejected for
planets with $9~M_{\oplus} \lesssim M_{\text{p}} \lesssim 20~M_{\oplus}$.
The occurrence-weighted median $M_{\text{p}} \approx 1~M_{\oplus}$
planet is less dense than the terrestrial planets in our solar system,
but the occurrence-weighted median planet with $15~M_{\oplus} \lesssim
M_{\text{p}} \lesssim 20~M_{\oplus}$ is denser than the ice giants.
The accretion of a 1\% H/He envelope by a planet with $M_{\text{p}}
\approx 2~M_{\oplus}$ is the expected outcome of the core-accretion
process in protoplanetary disks with a range of Shakura-Sunyaev $\alpha$
parameters and mass accretion rates.  The retention of that atmosphere
over Gyr timescales indicates that EUV-driven photoevaporation atmosphere
loss is less efficient than usually assumed and/or that core-driven
atmosphere loss is avoided because of large Bond albedos.  While Uranus
and Neptune have about 10\% of their masses in H/He atmospheres, the
observation that the occurrence-weighted median planet with $15~M_{\oplus}
\lesssim M_{\text{p}} \lesssim 20~M_{\oplus}$ has at most 5\% of its mass
in a H/He atmosphere implies that it formed in a hotter or shorter-lived
part of its parent protoplanetary disk than Uranus and Neptune.  It could
also indicate that giant impacts during disk dissipation are a common
stage of the formation of short-period planets with $10~M_{\oplus}
\lesssim M_{\text{p}} \lesssim 20~M_{\oplus}$, especially those in the
radius range $2~R_{\oplus} \lesssim R_{\text{p}} \lesssim 3~R_{\oplus}$.

\begin{acknowledgments}
We are grateful to the anonymous referee for an insightful review of our
paper and for suggestions that significantly improved our analyses.
We thank Bertram Bitsch, Eric Ford, Dan Fabrycky, Eve Lee, Eric
Lopez, Christoph Mordasini, and Leslie Rogers for helpful comments.
This research has made use of the NASA Exoplanet Archive, which is
operated by the California Institute of Technology, under contract
with the National Aeronautics and Space Administration under the
Exoplanet Exploration Program.  This research has made use of the
SIMBAD database, operated at CDS, Strasbourg, France \citep{wen00}.
This work has made use of data from the European Space Agency (ESA)
mission {\it Gaia} (\url{https://www.cosmos.esa.int/gaia}), processed
by the {\it Gaia} Data Processing and Analysis Consortium (DPAC,
\url{https://www.cosmos.esa.int/web/gaia/dpac/consortium}).  Funding for
the DPAC has been provided by national institutions, in particular the
institutions participating in the {\it Gaia} Multilateral Agreement.
This research has made use of the VizieR catalogue access tool,
CDS, Strasbourg, France (DOI: 10.26093/cds/vizier).  The original
description of the VizieR service was published in 2000, A\&AS 143,
23 \citep{och00}.  This research has made use of NASA's Astrophysics
Data System.  \end{acknowledgments}

\vspace{5mm}
\facilities{Exoplanet Archive}

\software{\texttt{R} \citep{r20},
          \texttt{TOPCAT} \citep{tay05}}

\clearpage
\bibliography{ms}{}
\bibliographystyle{aasjournal}

\clearpage
\begin{longrotatetable}
\begin{deluxetable*}{lclDDDDlll}
\tabletypesize{\tiny}
\tablecaption{Periods, Masses, Radii, and Occurrences for Transiting
Exoplanets with $M_{\text{p}} \lesssim 20~M_{\oplus}$\label{tbl-1}}
\tablewidth{0pt}
\tablehead{
\colhead{Host Star} &
\colhead{Gaia DR2} &
\colhead{Exoplanet} &
\twocolhead{Period} &
\twocolhead{Mass} &
\twocolhead{Radius} &
\twocolhead{Occurrence} &
\colhead{Mass} &
\colhead{Discovery} &
\colhead{Parameter} \\
\colhead{Simbad Name} &
\colhead{Source ID} &
\colhead{Name} &
\twocolhead{} &
\twocolhead{} &
\twocolhead{} &
\twocolhead{} &
\colhead{Method} &
\colhead{Reference} &
\colhead{Reference} \\
\colhead{} &
\colhead{} &
\colhead{} &
\twocolhead{(days)} &
\twocolhead{($M_{\oplus}$)} &
\twocolhead{($R_{\oplus}$)} &
\twocolhead{} &
\colhead{} &
\colhead{} &
\colhead{}}
\decimals
\startdata
Kepler-131 & 2102110174677569664 & Kepler-131 b & 16.092 &  16.1_{-  3.5}^{+  3.5} &  2.41_{-0.20}^{+0.20} & 0.036 & Doppler & \citet{mar14} & \citet{mar14}\\
Kepler-131 & 2102110174677569664 & Kepler-131 c & 25.5169 &   8.2_{-  5.9}^{+  5.9} &  0.84_{-0.07}^{+0.07} & 0.094 & Doppler & \citet{mar14} & \citet{mar14}\\
BD+41  3306 & 2101486923385239808 & Kepler-444 d & 6.189 &   0.2_{-  0.1}^{+  0.5} &  0.75_{-0.02}^{+0.02} & 0.061 & TTV & \citet{cam15} & \citet{had17}\\
BD+41  3306 & 2101486923385239808 & Kepler-444 e & 7.743 &   0.1_{-  0.1}^{+  0.2} &  0.73_{-0.02}^{+0.02} & 0.065 & TTV & \citet{cam15} & \citet{had17}\\
Kepler-10 & 2132155017099178624 & Kepler-10 b & 0.837491 &   4.6_{-  1.5}^{+  1.3} &  1.48_{-0.03}^{+0.05} & 0.0013 & Doppler & \citet{bat11} & \citet{est15}\\
Kepler-10 & 2132155017099178624 & Kepler-10 c & 45.2946 &   7.4_{-  1.2}^{+  1.3} &  2.62_{-0.03}^{+0.05} & 0.06 & Doppler & \citet{fre11} & \citet{raj17}\\
Kepler-106 & 2082074942519913344 & Kepler-106 c & 13.5708 &  10.4_{-  3.2}^{+  3.2} &  2.50_{-0.32}^{+0.32} & 0.031 & Doppler & \citet{mar14} & \citet{mar14}\\
Kepler-106 & 2082074942519913344 & Kepler-106 e & 43.8445 &  11.2_{-  5.8}^{+  5.8} &  2.56_{-0.33}^{+0.33} & 0.057 & Doppler & \citet{mar14} & \citet{mar14}\\
Kepler-289 & 2078515170549178880 & Kepler-289 b & 34.545 &   7.3_{-  6.8}^{+  6.8} &  2.15_{-0.10}^{+0.10} & 0.042 & TTV & \citet{row14} & \citet{sch14}\\
Kepler-289 & 2078515170549178880 & Kepler-289 d & 66.0634 &   4.0_{-  0.9}^{+  0.9} &  2.68_{-0.17}^{+0.17} & 0.056 & TTV & \citet{sch14} & \citet{sch14}\\
K2-263 & 657756997089784960 & K2-263 b & 50.818947 &  14.8_{-  3.1}^{+  3.1} &  2.41_{-0.12}^{+0.12} & 0.055 & Doppler & \citet{mor18} & \citet{mor18}\\
Kepler-538 & 2087171453788422528 & Kepler-538 b & 81.73778 &  10.6_{-  2.4}^{+  2.5} &  2.21_{-0.03}^{+0.04} & 0.055 & Doppler & \citet{mor16} & \citet{may19}\\
Kepler-549 & 2107622335702262400 & Kepler-549 b & 42.95 &  11.0_{-  3.2}^{+  4.2} &  2.87_{-0.08}^{+0.08} & 0.054 & TTV & \citet{mor16} & \citet{had17}\\
Kepler-310 & 2127798442794047104 & Kepler-310 d & 92.874 &   7.0_{-  4.1}^{+  3.4} &  2.44_{-0.05}^{+0.06} & 0.053 & TTV & \citet{row14} & \citet{had17}\\
HD  21749 & 4673947174316727040 & GJ 143 b & 35.61253 &  22.7_{-  1.9}^{+  2.2} &  2.61_{-0.16}^{+0.17} & 0.052 & Doppler & \citet{tri19} & \citet{dra19}\\
TYC  243-1528-1 & 3850421005290172416 & TOI-561 b & 0.446578 &   1.6_{-  0.4}^{+  0.4} &  1.42_{-0.07}^{+0.07} & 0.0014 & Doppler & \citet{lac21} & \citet{lac21}\\
TYC  243-1528-1 & 3850421005290172416 & TOI-561 c & 10.779 &   5.4_{-  1.0}^{+  1.0} &  2.88_{-0.10}^{+0.10} & 0.02 & Doppler & \citet{lac21} & \citet{lac21}\\
TYC  243-1528-1 & 3850421005290172416 & TOI-561 d & 25.62 &  11.9_{-  1.3}^{+  1.3} &  2.53_{-0.13}^{+0.13} & 0.047 & Doppler & \citet{lac21} & \citet{lac21}\\
TYC  243-1528-1 & 3850421005290172416 & TOI-561 e & 77.23 &  16.0_{-  2.3}^{+  2.3} &  2.67_{-0.11}^{+0.11} & 0.051 & Doppler & \citet{lac21} & \citet{lac21}\\
TYC  243-1528-1 & 3850421005290172416 & TOI-561 f & 16.287 &   3.0_{-  1.9}^{+  2.4} &  2.32_{-0.16}^{+0.16} & 0.038 & Doppler & \citet{wei21} & \citet{wei21}\\
BD+20   594 & 58200934326315136 & BD+20 594 b & 41.6855 &  16.3_{-  6.1}^{+  6.0} &  2.23_{-0.11}^{+0.14} & 0.049 & Doppler & \citet{esp16} & \citet{esp16}\\
Kepler-20 & 2102548708017562112 & Kepler-20 b & 3.69611525 &   9.7_{-  1.4}^{+  1.4} &  1.87_{-0.03}^{+0.07} & 0.0039 & Doppler & \citet{gau12} & \citet{buc16}\\
Kepler-20 & 2102548708017562112 & Kepler-20 c & 10.85409089 &  12.8_{-  2.2}^{+  2.2} &  3.05_{-0.06}^{+0.06} & 0.017 & Doppler & \citet{gau12} & \citet{buc16}\\
Kepler-20 & 2102548708017562112 & Kepler-20 d & 77.61130017 &  10.1_{-  3.7}^{+  4.0} &  2.74_{-0.06}^{+0.07} & 0.049 & Doppler & \citet{gau12} & \citet{buc16}\\
Kepler-359 & 2077683871034907648 & Kepler-359 c & 57.693 &   2.9_{-  1.9}^{+  2.4} &  3.61_{-0.15}^{+0.18} & 0.022 & TTV & \citet{row14} & \citet{had17}\\
Kepler-359 & 2077683871034907648 & Kepler-359 d & 77.083 &   2.7_{-  1.5}^{+  2.5} &  2.76_{-0.17}^{+0.18} & 0.049 & TTV & \citet{row14} & \citet{had17}\\
* nu.02 Lup & 5902750168276592256 & HD 136352 b & 11.57779 &   4.6_{-  0.4}^{+  0.4} &  1.48_{-0.06}^{+0.06} & 0.027 & Doppler & \citet{udr19} & \citet{kan20}\\
* nu.02 Lup & 5902750168276592256 & HD 136352 c & 27.5909 &  11.3_{-  0.7}^{+  0.7} &  2.61_{-0.08}^{+0.08} & 0.046 & Doppler & \citet{udr19} & \citet{kan20}\\
K2-138 & 2413596935442139520 & K2-138 b & 2.35309 &   3.1_{-  1.1}^{+  1.1} &  1.51_{-0.08}^{+0.11} & 0.0048 & Doppler & \citet{chr18} & \citet{lop19}\\
K2-138 & 2413596935442139520 & K2-138 c & 3.56004 &   6.3_{-  1.2}^{+  1.1} &  2.30_{-0.09}^{+0.12} & 0.0038 & Doppler & \citet{chr18} & \citet{lop19}\\
K2-138 & 2413596935442139520 & K2-138 d & 5.40479 &   7.9_{-  1.4}^{+  1.4} &  2.39_{-0.08}^{+0.10} & 0.0088 & Doppler & \citet{chr18} & \citet{lop19}\\
K2-138 & 2413596935442139520 & K2-138 e & 8.26146 &  13.0_{-  2.0}^{+  2.0} &  3.39_{-0.11}^{+0.16} & 0.0064 & Doppler & \citet{chr18} & \citet{lop19}\\
K2-138 & 2413596935442139520 & K2-138 f & 12.75758 &   1.6_{-  1.2}^{+  2.1} &  2.90_{-0.11}^{+0.16} & 0.023 & Doppler & \citet{chr18} & \citet{lop19}\\
K2-138 & 2413596935442139520 & K2-138 g & 41.96797 &   4.3_{-  3.0}^{+  5.3} &  3.01_{-0.25}^{+0.30} & 0.046 & Doppler & \citet{lop19} & \citet{lop19}\\
Kepler-595 & 2135237601028549888 & Kepler-595 b & 25.3029092 &  17.4_{-  3.8}^{+  7.1} &  3.71_{-0.01}^{+0.01} & 0.018 & TTV & \citet{yof21} & \citet{yof21}\\
Kepler-595 & 2135237601028549888 & Kepler-595 c & 12.38602 &   3.3_{-  1.0}^{+  1.7} &  1.01_{-0.02}^{+0.02} & 0.043 & TTV & \citet{yof21} & \citet{yof21}\\
HD  23472 & 4674216245427964416 & HD 23472 b & 17.667 &  17.9_{- 14.0}^{+  1.4} &  1.87_{-1.32}^{+1.32} & 0.0095 & Doppler & \citet{tri19} & \citet{tri19}\\
HD  23472 & 4674216245427964416 & HD 23472 c & 29.625 &  17.2_{- 13.8}^{+  1.1} &  2.15_{-0.34}^{+0.34} & 0.042 & Doppler & \citet{tri19} & \citet{tri19}\\
Kepler-11 & 2076960598545789824 & Kepler-11 d & 22.687 &   6.8_{-  0.8}^{+  0.7} &  4.04_{-0.40}^{+0.42} & 0.014 & TTV & \citet{lis11a} & \citet{had17}\\
Kepler-11 & 2076960598545789824 & Kepler-11 e & 31.995 &   6.7_{-  1.0}^{+  1.2} &  4.81_{-0.50}^{+0.55} & 0.0067 & TTV & \citet{lis11a} & \citet{had17}\\
Kepler-11 & 2076960598545789824 & Kepler-11 f & 46.686 &   1.7_{-  0.4}^{+  0.5} &  3.15_{-0.33}^{+0.38} & 0.041 & TTV & \citet{lis11a} & \citet{had17}\\
Kepler-30 & 2100243616248608896 & Kepler-30 b & 29.323 &   8.8_{-  0.5}^{+  0.6} &  2.13_{-0.13}^{+0.14} & 0.04 & TTV & \citet{san12} & \citet{had17}\\
HD  15337 & 5068777809824976256 & HD 15337 b & 4.75615 &   7.5_{-  1.0}^{+  1.1} &  1.64_{-0.06}^{+0.06} & 0.01 & Doppler & \citet{gan19} & \citet{gan19}\\
HD  15337 & 5068777809824976256 & HD 15337 c & 17.1784 &   8.1_{-  1.7}^{+  1.8} &  2.39_{-0.12}^{+0.12} & 0.039 & Doppler & \citet{gan19} & \citet{gan19}\\
HD   3167 & 2554032474712538880 & HD 3167 b & 0.96 &   5.6_{-  1.0}^{+  1.0} &  1.63_{-0.05}^{+0.05} & 0.00087 & Doppler & \citet{vand16} & \citet{dai19}\\
HD   3167 & 2554032474712538880 & HD 3167 c & 29.8454 &   9.8_{-  1.2}^{+  1.3} &  3.01_{-0.28}^{+0.42} & 0.037 & Doppler & \citet{vand16} & \citet{chr17}\\
Kepler-102 & 2119583201145735808 & Kepler-102 d & 10.3117 &   3.8_{-  1.8}^{+  1.8} &  1.18_{-0.04}^{+0.04} & 0.031 & Doppler & \citet{mar14} & \citet{mar14}\\
Kepler-102 & 2119583201145735808 & Kepler-102 e & 16.1457 &   8.9_{-  2.0}^{+  2.0} &  2.22_{-0.07}^{+0.07} & 0.037 & Doppler & \citet{wan14} & \citet{mar14}\\
BD+48  2893 & 2129550445852902656 & Kepler-68 b & 5.3988 &   7.7_{-  1.3}^{+  1.4} &  2.60_{-0.04}^{+0.04} & 0.0089 & Doppler & \citet{gil13} & \citet{mil19}\\
BD+48  2893 & 2129550445852902656 & Kepler-68 c & 9.6051 &   2.0_{-  1.8}^{+  1.7} &  1.04_{-0.02}^{+0.02} & 0.035 & Doppler & \citet{gil13} & \citet{mil19}\\
HD 119130 & 3616931735377523712 & K2-292 b & 16.9841 &  24.5_{-  4.4}^{+  4.4} &  2.63_{-0.10}^{+0.11} & 0.034 & Doppler & \citet{luq19} & \citet{luq19}\\
Kepler-79 & 2076085318570299136 & Kepler-79 b & 13.484542 &  12.5_{-  3.6}^{+  4.5} &  3.34_{-0.03}^{+0.03} & 0.013 & TTV & \citet{yof21} & \citet{yof21}\\
Kepler-79 & 2076085318570299136 & Kepler-79 c & 27.4023 &   9.5_{-  2.1}^{+  2.3} &  3.55_{-0.03}^{+0.03} & 0.02 & TTV & \citet{yof21} & \citet{yof21}\\
Kepler-79 & 2076085318570299136 & Kepler-79 d & 52.090767 &  11.3_{-  2.2}^{+  2.2} &  6.91_{-0.01}^{+0.01} & 0.005 & TTV & \citet{yof21} & \citet{yof21}\\
Kepler-79 & 2076085318570299136 & Kepler-79 e & 81.064 &   3.4_{-  0.8}^{+  1.0} &  3.40_{-0.13}^{+0.12} & 0.033 & TTV & \citet{row14} & \citet{had17}\\
Kepler-48 & 2075112109039378688 & Kepler-48 b & 4.778 &   3.9_{-  2.1}^{+  2.1} &  1.88_{-0.10}^{+0.10} & 0.0053 & Doppler & \citet{ste13} & \citet{mar14}\\
Kepler-48 & 2075112109039378688 & Kepler-48 c & 9.67395 &  14.6_{-  2.3}^{+  2.3} &  2.71_{-0.14}^{+0.14} & 0.021 & Doppler & \citet{ste13} & \citet{mar14}\\
Kepler-48 & 2075112109039378688 & Kepler-48 d & 42.8961 &   7.9_{-  4.6}^{+  4.6} &  2.04_{-0.11}^{+0.11} & 0.033 & Doppler & \citet{mar14} & \citet{mar14}\\
HD   5278 & 4617759514501503616 & HD 5278 b & 14.339156 &   7.8_{-  1.4}^{+  1.5} &  2.45_{-0.05}^{+0.05} & 0.033 & Doppler & \citet{soz21} & \citet{soz21}\\
Kepler-411 & 2132768952604988672 & Kepler-411 b & 3.005156 &  25.6_{-  2.6}^{+  2.6} &  2.40_{-0.05}^{+0.05} & 0.002 & TTV & \citet{wan14} & \citet{sun19}\\
Kepler-411 & 2132768952604988672 & Kepler-411 c & 7.834435 &  26.4_{-  5.9}^{+  5.9} &  4.42_{-0.06}^{+0.06} & 0.003 & TTV & \citet{mor16} & \citet{sun19}\\
Kepler-411 & 2132768952604988672 & Kepler-411 d & 58.02035 &  15.2_{-  5.1}^{+  5.1} &  3.32_{-0.10}^{+0.10} & 0.033 & TTV & \citet{sun19} & \citet{sun19}\\
Kepler-29 & 2086435189017387264 & Kepler-29 b & 10.33974 &   5.0_{-  1.3}^{+  1.5} &  2.55_{-0.12}^{+0.12} & 0.023 & TTV & \citet{fab12} & \citet{vis20}\\
Kepler-29 & 2086435189017387264 & Kepler-29 c & 13.28613 &   4.5_{-  1.1}^{+  1.1} &  2.34_{-0.11}^{+0.12} & 0.032 & TTV & \citet{fab12} & \citet{vis20}\\
Kepler-450 & 2135590578620605568 & Kepler-450 c & 15.4131395 &  12.5_{-  2.6}^{+  3.2} &  2.60_{-0.00}^{+0.00} & 0.032 & TTV & \citet{vane15} & \citet{yof21}\\
Kepler-96 & 2073731161099713408 & Kepler-96 b & 16.2385 &   8.5_{-  3.4}^{+  3.4} &  2.67_{-0.22}^{+0.22} & 0.032 & Doppler & \citet{mar14} & \citet{mar14}\\
K2-106 & 2582617711154563968 & EPIC 220674823 b & 0.571 &   7.7_{-  0.8}^{+  0.8} &  1.71_{-0.07}^{+0.07} & 0.00064 & Doppler & \citet{ada17} & \citet{dai19}\\
K2-106 & 2582617711154563968 & EPIC 220674823 c & 13.3397 &   5.8_{-  3.0}^{+  3.3} &  2.50_{-0.26}^{+0.27} & 0.031 & Doppler & \citet{ada17} & \citet{gue17}\\
K2-110 & 3613175223837135616 & K2-110 b & 13.86375 &  16.7_{-  3.2}^{+  3.2} &  2.59_{-0.10}^{+0.10} & 0.03 & Doppler & \citet{osb17} & \citet{osb17}\\
TOI-125 & 4698692744355471616 & TOI-125 b & 4.65382 &   9.5_{-  0.9}^{+  0.9} &  2.73_{-0.08}^{+0.08} & 0.0064 & Doppler & \citet{qui19} & \citet{nie20}\\
TOI-125 & 4698692744355471616 & TOI-125 c & 9.15059 &   6.6_{-  1.0}^{+  1.0} &  2.76_{-0.10}^{+0.10} & 0.019 & Doppler & \citet{qui19} & \citet{nie20}\\
TOI-125 & 4698692744355471616 & TOI-125 d & 19.98 &  13.6_{-  1.2}^{+  1.2} &  2.93_{-0.17}^{+0.17} & 0.03 & Doppler & \citet{nie20} & \citet{nie20}\\
Kepler-305 & 2073547954981763328 & Kepler-305 d & 16.739 &   9.1_{-  3.8}^{+  6.1} &  2.80_{-0.09}^{+0.09} & 0.03 & TTV & \citet{row14} & \citet{had17}\\
Kepler-1655 & 2100392428275714688 & Kepler-1655 b & 11.8728787 &   5.0_{-  2.8}^{+  3.1} &  2.21_{-0.08}^{+0.08} & 0.029 & Doppler & \citet{hay18} & \citet{hay18}\\
Kepler-85 & 2127398083121764736 & Kepler-85 e & 25.215 &   0.6_{-  0.4}^{+  0.5} &  1.28_{-0.09}^{+0.09} & 0.028 & TTV & \citet{row14} & \citet{had17}\\
CD-39  7945 & 6140553127216043648 & TOI-763 b & 5.6057 &   9.8_{-  0.8}^{+  0.8} &  2.28_{-0.11}^{+0.11} & 0.0093 & Doppler & \citet{fri20} & \citet{fri20}\\
CD-39  7945 & 6140553127216043648 & TOI-763 c & 12.2737 &   9.3_{-  1.0}^{+  1.0} &  2.63_{-0.12}^{+0.12} & 0.028 & Doppler & \citet{fri20} & \citet{fri20}\\
Kepler-307 & 2073691303802278784 & Kepler-307 b & 10.416 &   8.8_{-  0.9}^{+  0.9} &  3.04_{-0.06}^{+0.07} & 0.016 & TTV & \citet{xie14} & \citet{had17}\\
Kepler-307 & 2073691303802278784 & Kepler-307 c & 13.084 &   3.9_{-  0.7}^{+  0.7} &  2.73_{-0.07}^{+0.06} & 0.028 & TTV & \citet{xie14} & \citet{had17}\\
Kepler-406 & 2126601108991279872 & Kepler-406 b & 2.42629 &   6.3_{-  1.4}^{+  1.4} &  1.43_{-0.03}^{+0.03} & 0.0059 & Doppler & \citet{mar14} & \citet{mar14}\\
Kepler-406 & 2126601108991279872 & Kepler-406 c & 4.62332 &   2.7_{-  1.8}^{+  1.8} &  0.85_{-0.03}^{+0.03} & 0.027 & Doppler & \citet{mar14} & \citet{mar14}\\
Kepler-107 & 2086625752425381632 & Kepler-107 b & 3.1800218 &   3.5_{-  1.5}^{+  1.5} &  1.54_{-0.02}^{+0.02} & 0.0074 & Doppler & \citet{row14} & \citet{bon19}\\
Kepler-107 & 2086625752425381632 & Kepler-107 c & 4.901452 &   9.4_{-  1.8}^{+  1.8} &  1.60_{-0.03}^{+0.03} & 0.011 & Doppler & \citet{row14} & \citet{bon19}\\
Kepler-107 & 2086625752425381632 & Kepler-107 e & 14.749143 &   8.6_{-  3.6}^{+  3.6} &  2.90_{-0.04}^{+0.04} & 0.025 & Doppler & \citet{row14} & \citet{bon19}\\
Kepler-454 & 2099541715513813504 & Kepler-454 b & 10.57375339 &   6.8_{-  1.4}^{+  1.4} &  2.37_{-0.13}^{+0.13} & 0.025 & Doppler & \citet{get16} & \citet{get16}\\
K2-38 & 6237129658760381056 & K2-38 b & 4.01593 &  12.0_{-  2.9}^{+  2.9} &  1.55_{-0.16}^{+0.16} & 0.0097 & Doppler & \citet{sin16} & \citet{sin16}\\
K2-38 & 6237129658760381056 & K2-38 c & 10.56103 &   9.9_{-  4.6}^{+  4.6} &  2.42_{-0.29}^{+0.29} & 0.025 & Doppler & \citet{sin16} & \citet{sin16}\\
Kepler-51 & 2135275362382289280 & Kepler-51 b & 45.155 &   2.3_{-  1.6}^{+  1.7} &  7.47_{-0.16}^{+0.09} & 0.0044 & TTV & \citet{ste13} & \citet{had17}\\
Kepler-51 & 2135275362382289280 & Kepler-51 c & 85.316 &   3.9_{-  0.8}^{+  0.8} &  4.13_{-0.11}^{+0.10} & 0.024 & TTV & \citet{ste13} & \citet{had17}\\
Kepler-51 & 2135275362382289280 & Kepler-51 d & 130.18 &   6.2_{-  1.5}^{+  1.6} & 10.21_{-0.23}^{+0.12} & 0.013 & TTV & \citet{mas14} & \citet{had17}\\
K2-32 & 4130539180358512768 & K2-32 b & 8.99213 &  16.5_{-  2.7}^{+  2.7} &  5.13_{-0.28}^{+0.28} & 0.0024 & Doppler & \citet{dai16} & \citet{pet17}\\
K2-32 & 4130539180358512768 & K2-32 d & 31.7154 &  10.3_{-  4.7}^{+  4.7} &  3.43_{-0.35}^{+0.35} & 0.023 & Doppler & \citet{dai16} & \citet{pet17}\\
Kepler-100 & 2101733244046205568 & Kepler-100 b & 6.88705 &   7.3_{-  3.2}^{+  3.2} &  1.32_{-0.04}^{+0.04} & 0.022 & Doppler & \citet{mar14} & \citet{mar14}\\
HD  97658 & 3997075206232885888 & HD 97658 b & 9.4903 &   7.5_{-  0.8}^{+  0.8} &  2.25_{-0.10}^{+0.10} & 0.022 & Doppler & \citet{how11} & \citet{van14}\\
HD 106315 & 3698307419878650240 & HD 106315 b & 9.55237 &  12.6_{-  3.2}^{+  3.2} &  2.44_{-0.17}^{+0.17} & 0.022 & Doppler & \citet{cro17} & \citet{bar17}\\
HD 106315 & 3698307419878650240 & HD 106315 c & 21.05704 &  15.2_{-  3.7}^{+  3.7} &  4.35_{-0.23}^{+0.23} & 0.01 & Doppler & \citet{cro17} & \citet{bar17}\\
Kepler-279 & 2102456726998624640 & Kepler-279 c & 35.735 &   7.4_{-  1.5}^{+  2.2} &  4.25_{-0.13}^{+0.13} & 0.013 & TTV & \citet{xie14} & \citet{had17}\\
Kepler-279 & 2102456726998624640 & Kepler-279 d & 54.414 &   4.5_{-  0.9}^{+  1.2} &  3.71_{-0.11}^{+0.12} & 0.02 & TTV & \citet{xie14} & \citet{had17}\\
Kepler-19 & 2051106987063242880 & Kepler-19 b & 9.28716 &   8.4_{-  1.5}^{+  1.6} &  2.21_{-0.05}^{+0.05} & 0.02 & joint & \citet{bal11} & \citet{mal17}\\
Kepler-1662 & 2127790368255394048 & KOI-1783.02 & 284.215 &  15.0_{-  3.6}^{+  4.3} &  5.44_{-0.30}^{+0.52} & 0.02 & TTV & \citet{vis20} & \citet{vis20}\\
BD-00  4534 & 2645940376800212096 & HIP 116454 b & 9.1205 &  11.8_{-  1.3}^{+  1.3} &  2.53_{-0.18}^{+0.18} & 0.02 & Doppler & \citet{vand15} & \citet{vand15}\\
Kepler-60 & 2102316985943863424 & Kepler-60 b & 7.133 &   3.7_{-  0.6}^{+  0.6} &  1.98_{-0.05}^{+0.06} & 0.0098 & TTV & \citet{ste13} & \citet{had17}\\
Kepler-60 & 2102316985943863424 & Kepler-60 c & 8.919 &   2.0_{-  0.5}^{+  0.3} &  2.21_{-0.06}^{+0.06} & 0.019 & TTV & \citet{ste13} & \citet{had17}\\
Kepler-60 & 2102316985943863424 & Kepler-60 d & 11.899 &   3.9_{-  0.6}^{+  0.7} &  1.89_{-0.10}^{+0.09} & 0.013 & TTV & \citet{ste13} & \citet{had17}\\
HD 219134 & 2009481748875806976 & HD 219134 b & 3.092926 &   4.7_{-  0.2}^{+  0.2} &  1.60_{-0.06}^{+0.06} & 0.0062 & Doppler & \citet{mot15} & \citet{gil17}\\
HD 219134 & 2009481748875806976 & HD 219134 c & 6.76458 &   4.4_{-  0.2}^{+  0.2} &  1.51_{-0.05}^{+0.05} & 0.017 & Doppler & \citet{mot15} & \citet{gil17}\\
CD-61  1276 & 5481210874877547904 & TOI-220 b & 10.695264 &  13.8_{-  1.0}^{+  1.0} &  3.03_{-0.15}^{+0.15} & 0.017 & Doppler & \citet{hoy21} & \citet{hoy21}\\
BD+40  3638 & 2102119176929154304 & Kepler-65 b & 2.1549209 &   2.4_{-  1.6}^{+  2.4} &  1.44_{-0.03}^{+0.04} & 0.0046 & joint & \citet{cha13} & \citet{mil19}\\
BD+40  3638 & 2102119176929154304 & Kepler-65 c & 5.859697 &   5.4_{-  1.7}^{+  1.7} &  2.62_{-0.06}^{+0.07} & 0.01 & joint & \citet{cha13} & \citet{mil19}\\
BD+40  3638 & 2102119176929154304 & Kepler-65 d & 8.13167 &   4.1_{-  0.8}^{+  0.8} &  1.59_{-0.04}^{+0.04} & 0.017 & joint & \citet{cha13} & \citet{mil19}\\
Kepler-177 & 2106339927188171776 & Kepler-177 b & 36.855 &   5.4_{-  0.9}^{+  1.0} &  3.92_{-0.61}^{+0.72} & 0.016 & TTV & \citet{xie14} & \citet{had17}\\
Kepler-177 & 2106339927188171776 & Kepler-177 c & 49.412 &  13.5_{-  2.8}^{+  2.6} &  8.36_{-1.34}^{+1.37} & 0.0041 & TTV & \citet{xie14} & \citet{had17}\\
Kepler-36 & 2129931456691176576 & Kepler-36 b & 13.848 &   3.9_{-  0.2}^{+  0.2} &  1.71_{-0.04}^{+0.04} & 0.016 & TTV & \citet{car12} & \citet{had17}\\
Kepler-36 & 2129931456691176576 & Kepler-36 c & 16.233 &   7.5_{-  0.3}^{+  0.3} &  4.15_{-0.07}^{+0.07} & 0.0084 & TTV & \citet{car12} & \citet{had17}\\
K2-314 & 6259263137059042048 & EPIC 249893012 b & 3.5951 &   8.8_{-  1.1}^{+  1.1} &  1.95_{-0.08}^{+0.09} & 0.0038 & Doppler & \citet{hid20} & \citet{hid20}\\
K2-314 & 6259263137059042048 & EPIC 249893012 c & 15.624 &  14.7_{-  1.9}^{+  1.8} &  3.67_{-0.14}^{+0.17} & 0.01 & Doppler & \citet{hid20} & \citet{hid20}\\
K2-314 & 6259263137059042048 & EPIC 249893012 d & 35.747 &  10.2_{-  2.4}^{+  2.5} &  3.94_{-0.12}^{+0.13} & 0.016 & Doppler & \citet{hid20} & \citet{hid20}\\
Kepler-33 & 2127355923723254272 & Kepler-33 d & 21.776 &   4.1_{-  2.0}^{+  1.7} &  5.37_{-0.13}^{+0.12} & 0.004 & TTV & \citet{lis12} & \citet{had17}\\
Kepler-33 & 2127355923723254272 & Kepler-33 e & 31.784 &   5.5_{-  1.1}^{+  1.2} &  4.00_{-0.10}^{+0.11} & 0.015 & TTV & \citet{lis12} & \citet{had17}\\
Kepler-33 & 2127355923723254272 & Kepler-33 f & 41.028 &   9.6_{-  1.8}^{+  1.7} &  4.43_{-0.11}^{+0.10} & 0.011 & TTV & \citet{lis12} & \citet{had17}\\
Kepler-105 & 2130368310704421120 & Kepler-105 c & 7.126 &   3.6_{-  1.3}^{+  1.3} &  1.60_{-0.04}^{+0.04} & 0.015 & TTV & \citet{row14} & \citet{had17}\\
Kepler-103 & 2101243789577188736 & Kepler-103 b & 15.9654 &   9.7_{-  8.6}^{+  8.6} &  3.37_{-0.09}^{+0.09} & 0.015 & Doppler & \citet{mar14} & \citet{mar14}\\
Kepler-103 & 2101243789577188736 & Kepler-103 c & 179.612 &  36.1_{- 25.2}^{+ 25.2} &  5.14_{-0.14}^{+0.14} & 0.011 & Doppler & \citet{mar14} & \citet{mar14}\\
Kepler-82 & 2125850623586710400 & Kepler-82 b & 26.44 &  12.2_{-  0.9}^{+  1.0} &  4.07_{-0.10}^{+0.24} & 0.015 & TTV & \citet{xie13} & \citet{fre19}\\
Kepler-82 & 2125850623586710400 & Kepler-82 c & 51.54 &  13.9_{-  1.2}^{+  1.3} &  5.34_{-0.13}^{+0.32} & 0.0053 & TTV & \citet{xie13} & \citet{fre19}\\
Kepler-1659 & 2073779161653824640 & KOI-1599.01 & 20.4415 &   4.6_{-  0.3}^{+  0.3} &  1.90_{-0.30}^{+0.30} & 0.01 & TTV & \citet{pan19} & \citet{pan19}\\
Kepler-1659 & 2073779161653824640 & KOI-1599.02 & 13.6088 &   9.0_{-  0.3}^{+  0.3} &  1.90_{-0.20}^{+0.20} & 0.013 & TTV & \citet{pan19} & \citet{pan19}\\
BD+38  3583 & 2052747119115620352 & Kepler-93 b & 4.72673978 &   4.0_{-  0.7}^{+  0.7} &  1.48_{-0.02}^{+0.02} & 0.013 & Doppler & \citet{mar14} & \citet{dre15}\\
HD 183579 & 6641996571978861440 & HD 183579 b & 17.471278 &  19.7_{-  3.9}^{+  4.0} &  3.55_{-0.12}^{+0.15} & 0.013 & Doppler & \citet{pal21} & \citet{pal21}\\
K2-10 & 3798690151434945280 & K2-10 b & 19.3044 &  27.0_{- 16.0}^{+ 17.0} &  3.84_{-0.34}^{+0.35} & 0.013 & Doppler & \citet{mon15} & \citet{vane16}\\
Kepler-99 & 2076871091425583232 & Kepler-99 b & 4.60358 &   6.2_{-  1.3}^{+  1.3} &  1.48_{-0.08}^{+0.08} & 0.013 & Doppler & \citet{mar14} & \citet{mar14}\\
Kepler-25 & 2100451630105041152 & Kepler-25 b & 6.238297 &   8.7_{-  2.3}^{+  2.5} &  2.75_{-0.04}^{+0.04} & 0.012 & joint & \citet{ste12} & \citet{mil19}\\
Kepler-25 & 2100451630105041152 & Kepler-25 c & 12.7207 &  15.2_{-  1.6}^{+  1.3} &  5.22_{-0.06}^{+0.07} & 0.0033 & joint & \citet{ste12} & \citet{mil19}\\
Kepler-87 & 2086306584816171392 & Kepler-87 c & 191.2318 &   6.4_{-  0.8}^{+  0.8} &  6.14_{-0.29}^{+0.29} & 0.01 & TTV & \citet{ofi14} & \citet{ofi14}\\
~* pi. Men & 4623036865373793408 & pi Men c & 6.2679 &   4.8_{-  0.9}^{+  0.8} &  2.04_{-0.05}^{+0.05} & 0.009 & Doppler & \citet{gan18} & \citet{hua18}\\
Kepler-95 & 2106705269988803072 & Kepler-95 b & 11.5231 &  13.0_{-  2.9}^{+  2.9} &  3.42_{-0.09}^{+0.09} & 0.0087 & Doppler & \citet{mar14} & \citet{mar14}\\
BD-14  1137 & 2984582227215748864 & TOI-421 b & 5.19672 &   7.2_{-  0.7}^{+  0.7} &  2.68_{-0.18}^{+0.19} & 0.0082 & Doppler & \citet{car20} & \citet{car20}\\
BD-14  1137 & 2984582227215748864 & TOI-421 c & 16.06819 &  16.4_{-  1.0}^{+  1.1} &  5.09_{-0.15}^{+0.16} & 0.0038 & Doppler & \citet{car20} & \citet{car20}\\
K2-66 & 2614121364991061888 & K2-66 b & 5.06963 &  21.3_{-  3.6}^{+  3.6} &  2.49_{-0.24}^{+0.34} & 0.0078 & Doppler & \citet{cro16} & \citet{sin17}\\
Kepler-88 & 2101507367429089664 & KOI-142 b & 10.91647 &   9.5_{-  1.1}^{+  1.1} &  3.44_{-0.08}^{+0.08} & 0.0078 & joint & \citet{wei21} & \citet{wei21}\\
K2-111 & 53006669599267328 & K2-111 b & 5.3518 &   5.3_{-  0.8}^{+  0.8} &  1.82_{-0.09}^{+0.11} & 0.0074 & Doppler & \citet{fri17} & \citet{mor20}\\
Kepler-113 & 2133369840008452224 & Kepler-113 b & 4.754 &  11.7_{-  4.2}^{+  4.2} &  1.82_{-0.05}^{+0.05} & 0.0064 & Doppler & \citet{mar14} & \citet{mar14}\\
Kepler-89 & 2076970047474270208 & KOI-94 b & 3.743208 &  10.5_{-  4.6}^{+  4.6} &  1.71_{-0.16}^{+0.16} & 0.0064 & Doppler & \citet{wei13} & \citet{wei13}\\
Kepler-89 & 2076970047474270208 & KOI-94 c & 10.424 &   7.8_{-  2.4}^{+  3.0} &  4.32_{-0.09}^{+0.10} & 0.0044 & TTV & \citet{wei13} & \citet{had17}\\
Kepler-89 & 2076970047474270208 & KOI-94 e & 54.32031 &  35.0_{- 28.0}^{+ 18.0} &  6.56_{-0.62}^{+0.62} & 0.0052 & Doppler & \citet{wei13} & \citet{wei13}\\
Kepler-97 & 2131216137248338304 & Kepler-97 b & 2.58664 &   3.5_{-  1.9}^{+  1.9} &  1.48_{-0.13}^{+0.13} & 0.006 & Doppler & \citet{mar14} & \citet{mar14}\\
K2-36 & 3811989156889528320 & K2-36 b & 1.422614 &   3.9_{-  1.1}^{+  1.1} &  1.43_{-0.08}^{+0.08} & 0.0019 & Doppler & \citet{sin16} & \citet{dam19}\\
K2-36 & 3811989156889528320 & K2-36 c & 5.340888 &   7.8_{-  2.3}^{+  2.3} &  3.20_{-0.30}^{+0.30} & 0.0052 & Doppler & \citet{sin16} & \citet{dam19}\\
WASP-47 & 2613413008919918976 & WASP-47 d & 9.03077 &  13.1_{-  1.5}^{+  1.5} &  3.58_{-0.05}^{+0.05} & 0.005 & Doppler & \citet{bec15} & \citet{van17}\\
WASP-47 & 2613413008919918976 & WASP-47 e & 0.789592 &   6.8_{-  0.7}^{+  0.7} &  1.81_{-0.03}^{+0.03} & 0.00043 & Doppler & \citet{bec15} & \citet{van17}\\
HD  86226 & 5660492297395345408 & HD 86226 c & 3.98442 &   7.2_{-  1.1}^{+  1.2} &  2.16_{-0.08}^{+0.08} & 0.0048 & Doppler & \citet{tes20} & \citet{tes20}\\
HD 179070 & 2099606483621385216 & Kepler-21 b & 2.78578 &   5.1_{-  1.7}^{+  1.7} &  1.64_{-0.02}^{+0.02} & 0.0048 & Doppler & \citet{how12} & \citet{lop16}\\
K2-105 & 651907079835937280 & K2-105 b & 8.266902 &  30.0_{- 19.0}^{+ 19.0} &  3.59_{-0.39}^{+0.44} & 0.0045 & Doppler & \citet{nar17} & \citet{nar17}\\
K2-98 & 601199802584914816 & K2-98 b & 10.13675 &  32.2_{-  8.1}^{+  8.1} &  4.30_{-0.20}^{+0.30} & 0.0043 & Doppler & \citet{bar16} & \citet{bar16}\\
CD-24 12581 & 6049750234317822208 & K2-24 b & 20.88977 &  19.0_{-  2.1}^{+  2.2} &  5.40_{-0.20}^{+0.20} & 0.0039 & joint & \citet{pet16} & \citet{pet18}\\
CD-24 12581 & 6049750234317822208 & K2-24 c & 42.3391 &  15.4_{-  1.8}^{+  1.9} &  7.50_{-0.30}^{+0.30} & 0.0042 & joint & \citet{pet16} & \citet{pet18}\\
Kepler-223 & 2086337508581280256 & Kepler-223 b & 7.385 &   4.0_{-  2.1}^{+  1.7} &  3.94_{-0.48}^{+0.43} & 0.0035 & TTV & \citet{row14} & \citet{had17}\\
Kepler-223 & 2086337508581280256 & Kepler-223 c & 9.848 &  12.4_{-  2.9}^{+  2.8} &  4.34_{-0.48}^{+0.50} & 0.004 & TTV & \citet{row14} & \citet{had17}\\
Kepler-223 & 2086337508581280256 & Kepler-223 d & 14.787 &   5.9_{-  1.9}^{+  1.9} &  5.87_{-0.69}^{+0.61} & 0.0028 & TTV & \citet{row14} & \citet{had17}\\
K2-285 & 2657374606238804992 & K2-285 b & 3.471745 &   9.7_{-  1.4}^{+  1.2} &  2.59_{-0.06}^{+0.06} & 0.0028 & Doppler & \citet{pal19} & \citet{pal19}\\
K2-285 & 2657374606238804992 & K2-285 c & 7.138048 &  15.7_{-  2.1}^{+  2.3} &  3.53_{-0.08}^{+0.08} & 0.0039 & Doppler & \citet{pal19} & \citet{pal19}\\
K2-19 & 3798833775141351552 & K2-19 b & 7.9194 &  28.5_{-  5.0}^{+  5.4} &  7.74_{-0.39}^{+0.39} & 0.00089 & Doppler & \citet{dai16} & \citet{dai16}\\
K2-19 & 3798833775141351552 & K2-19 c & 11.90715 &  25.6_{-  7.1}^{+  7.1} &  4.86_{-0.44}^{+0.62} & 0.0038 & Doppler & \citet{dai16} & \citet{dai16}\\
HD 285181 & 3409148746676599168 & K2-291 b & 2.225177 &   6.5_{-  1.2}^{+  1.2} &  1.59_{-0.07}^{+0.10} & 0.0036 & Doppler & \citet{kos19} & \citet{kos19}\\
CoRoT-24 & 3105404467618982272 & CoRoT-24 c & 11.759 &  28.0_{- 11.0}^{+ 11.0} &  5.00_{-0.50}^{+0.50} & 0.0034 & Doppler & \citet{alo14} & \citet{alo14}\\
Kepler-46 & 2102700131386216576 & Kepler-46 b & 33.648 & 281.3_{-109.0}^{+118.9} &  7.99_{-0.19}^{+0.10} & 0.0033 & TTV & \citet{nes12} & \citet{saa17}\\
BD-15  6276 & 2597119620985658496 & K2-265 b & 2.369172 &   6.5_{-  0.8}^{+  0.8} &  1.71_{-0.11}^{+0.11} & 0.0031 & Doppler & \citet{lam18} & \citet{lam18}\\
BD+19  2158 & 636363799347569408 & EPIC 211945201 b & 19.49213 &  27.0_{- 12.6}^{+ 14.0} &  6.12_{-0.10}^{+0.10} & 0.003 & Doppler & \citet{cha18} & \citet{cha18}\\
CoRoT-22 & 4285572454508522496 & CoRoT-22 b & 9.75598 &  12.2_{-  8.8}^{+ 14.0} &  4.88_{-0.39}^{+0.17} & 0.003 & Doppler & \citet{mou14} & \citet{mou14}\\
TOI-216 & 4664811297844004352 & TOI-216.01 & 34.556 & 200.0_{-100.0}^{+170.0} & 11.29_{-0.42}^{+0.58} & 0.0029 & TTV & \citet{kip19} & \citet{kip19}\\
TOI-216 & 4664811297844004352 & TOI-216.02 & 17.089 &  30.0_{- 14.0}^{+ 20.0} &  7.69_{-0.83}^{+1.62} & 0.0018 & TTV & \citet{kip19} & \citet{kip19}\\
HD 158259 & 1416050859226670848 & HD 158259 b & 2.178 &   2.2_{-  0.4}^{+  0.4} &  1.20_{-1.30}^{+1.30} & 0.0028 & Doppler & \citet{har20} & \citet{har20}\\
Kepler-18 & 2079295583282164992 & Kepler-18 d & 14.859 &  14.9_{-  4.2}^{+  1.8} &  5.93_{-0.06}^{+0.13} & 0.0027 & TTV & \citet{coc11} & \citet{had17}\\
K2-27 & 3798552815560689792 & K2-27 b & 6.771315 &  30.9_{-  4.6}^{+  4.6} &  4.48_{-0.23}^{+0.23} & 0.0024 & Doppler & \citet{vane16} & \citet{pet17}\\
K2-216 & 2556231154370582400 & K2-216 b & 2.1748 &   8.0_{-  1.6}^{+  1.6} &  1.75_{-0.10}^{+0.17} & 0.0024 & Doppler & \citet{may18} & \citet{per18}\\
HD 110113 & 6133384959942131968 & HD 110113 b & 2.541 &   4.5_{-  0.6}^{+  0.6} &  2.05_{-0.12}^{+0.12} & 0.002 & Doppler & \citet{osb21} & \citet{osb21}\\
Kepler-117 & 2130959744881558272 & Kepler-117 b & 18.7959228 &  29.9_{- 10.5}^{+ 10.5} &  8.06_{-0.27}^{+0.27} & 0.002 & joint & \citet{row14} & \citet{bru15}\\
HD 219666 & 6492940453524576128 & HD 219666 b & 6.03607 &  16.6_{-  1.3}^{+  1.3} &  4.71_{-0.17}^{+0.17} & 0.0019 & Doppler & \citet{esp19} & \citet{esp19}\\
BD-05  3504 & 3583630929786305280 & K2-229 b & 0.584 &   2.5_{-  0.4}^{+  0.4} &  1.20_{-0.05}^{+0.04} & 0.0017 & Doppler & \citet{may18} & \citet{dai19}\\
Kepler-78 & 2078373642776670080 & Kepler-78 b & 0.355 &   1.8_{-  0.2}^{+  0.2} &  1.23_{-0.02}^{+0.02} & 0.0016 & Doppler & \citet{san13} & \citet{dai19}\\
TYC 4450-1307-1 & 2274696151198328960 & TOI-1444 b & 0.4702694 &   3.9_{-  0.7}^{+  0.7} &  1.40_{-0.06}^{+0.06} & 0.0015 & Doppler & \citet{dai21} & \citet{dai21}\\
StKM 1-2122 & 2643952940813536768 & K2-141 b & 0.2803244 &   5.1_{-  0.4}^{+  0.4} &  1.51_{-0.05}^{+0.05} & 0.0012 & Doppler & \citet{mal18} & \citet{mal18}\\
Kepler-4 & 2132152916856093952 & Kepler-4 b & 3.21346 &  24.5_{-  3.8}^{+  3.8} &  4.00_{-0.21}^{+0.21} & 0.0012 & Doppler & \citet{bor10} & \citet{bor10}\\
HD  89345 & 3877070590167177344 & HD 89345 b & 11.8143 &  35.0_{-  5.7}^{+  5.4} &  7.40_{-0.34}^{+0.31} & 0.0011 & Doppler & \citet{van18} & \citet{yu18}\\
HATS-46 & 4919771654727611008 & HATS-46 b & 4.7423729 &  55.0_{- 19.7}^{+ 19.7} & 10.12_{-0.50}^{+0.65} & 0.0011 & Doppler & \citet{bra18} & \citet{bra18}\\
CoRoT-32 & 3326541506073371776 & CoRoTID 223977153 b & 6.71837 &  47.7_{- 31.8}^{+ 31.8} &  6.39_{-0.56}^{+0.67} & 0.001 & Doppler & \citet{bou18} & \citet{bou18}\\
Kepler-94 & 2119602202081351168 & Kepler-94 b & 2.50806 &  10.8_{-  1.4}^{+  1.4} &  3.51_{-0.15}^{+0.15} & 0.00099 & Doppler & \citet{mar14} & \citet{mar14}\\
CoRoT-7 & 3107267177757848576 & CoRoT-7 b & 0.854 &   4.6_{-  1.2}^{+  1.1} &  1.58_{-0.10}^{+0.10} & 0.00099 & Doppler & \citet{leg09} & \citet{dai19}\\
BD-09  6003 & 2609222216055014144 & K2-39 b & 4.60497 &  39.8_{-  4.4}^{+  4.4} &  5.71_{-0.63}^{+0.63} & 0.00093 & Doppler & \citet{vane16} & \citet{pet17}\\
HD  80653 & 606477252238780160 & HD 80653 b & 0.719573 &   5.6_{-  0.4}^{+  0.4} &  1.61_{-0.07}^{+0.07} & 0.00087 & Doppler & \citet{fru20} & \citet{fru20}\\
Gaia DR2 6585082036193768832 & 6585082036193768832 & NGTS-14 A b & 3.5357173 &  29.2_{-  3.8}^{+  3.8} &  4.98_{-0.34}^{+0.34} & 0.00079 & Doppler & \citet{smi21} & \citet{smi21}\\
K2-131 & 3580920878437938688 & K2-131 b & 0.369 &   6.3_{-  1.4}^{+  1.4} &  1.65_{-0.06}^{+0.06} & 0.00077 & Doppler & \citet{dai17} & \citet{dai19}\\
HAT-P-26 & 3668036348641580288 & HAT-P-26 b & 4.234516 &  18.8_{-  2.2}^{+  2.2} &  6.33_{-0.36}^{+0.81} & 0.00074 & Doppler & \citet{har11} & \citet{har11}\\
CD-44 14956 & 6520880040423258240 & TOI-132 b & 2.1097019 &  22.4_{-  1.9}^{+  1.9} &  3.42_{-0.14}^{+0.13} & 0.00074 & Doppler & \citet{dia20} & \citet{dia20}\\
HATS-37 & 6194574671813047424 & HATS-37 A b & 4.3315366 &  31.5_{- 13.3}^{+ 13.3} &  6.79_{-0.18}^{+0.18} & 0.00066 & Doppler & \citet{jor20} & \citet{jor20}\\
HD 213885 & 6407428994690988928 & HD 213885 b & 1.008035 &   8.8_{-  0.6}^{+  0.7} &  1.75_{-0.05}^{+0.05} & 0.00065 & Doppler & \citet{esp20} & \citet{esp20}\\
TYC 6622-794-1 & 5472386851683941376 & HATS-38 b & 4.375021 &  23.5_{-  3.5}^{+  3.5} &  6.88_{-0.19}^{+0.19} & 0.00064 & Doppler & \citet{jor20} & \citet{jor20}\\
Kepler-98 & 2099862910348755712 & Kepler-98 b & 1.54168 &   3.5_{-  1.6}^{+  1.6} &  1.99_{-0.22}^{+0.22} & 0.00058 & Doppler & \citet{mar14} & \citet{mar14}\\
NGTS-4 & 2891248292906892032 & NGTS-4 b & 1.3373508 &  20.6_{-  3.0}^{+  3.0} &  3.18_{-0.26}^{+0.26} & 0.00041 & Doppler & \citet{wes19} & \citet{wes19}\\
TYC 8688-915-1 & 5880886001621564928 & TOI-824 b & 1.392978 &  18.5_{-  1.9}^{+  1.8} &  2.93_{-0.19}^{+0.20} & 0.00035 & Doppler & \citet{bur20} & \citet{bur20}\\
~* rho01 Cnc & 704967037090946688 & 55 Cnc e & 0.7365474 &   8.0_{-  0.3}^{+  0.3} &  1.88_{-0.03}^{+0.03} & 0.00027 & Doppler & \citet{mca04} & \citet{bou18}
\enddata
\tablecomments{This table is ordered in descending order of the maximum
occurrence of individual exoplanets in an exoplanet system.  It is
available in its entirety in a machine-readable format.}
\end{deluxetable*}
\end{longrotatetable}

\end{document}